\documentclass[fleqn,usenatbib]{mnras}

\usepackage{newtxtext,newtxmath}

\usepackage[T1]{fontenc}

\DeclareRobustCommand{\VAN}[3]{#2}
\let\VANthebibliography\thebibliography
\def\thebibliography{\DeclareRobustCommand{\VAN}[3]{##3}\VANthebibliography}
\defcitealias{Schnorr.et.al.2021}{SM21}
\defcitealias{Spiniello.etal.2021b}{S21b}

\usepackage{graphicx}	
\usepackage{amsmath}	

\usepackage{amssymb}	

\title[MCGs in SDSS]{Massive Compact Quiescent Galaxies in the $M_\star$ vs. $\sigma_\mathrm{e}$ Plane: Insights from stellar Population Properties}

\author[K. Slodkowski Clerici et al.]{
K. Slodkowski Clerici,$^{1}$\thanks{E-mail: katia.clerici@ufrgs.br}
A. Schnorr-Müller,$^{1}$
M. Trevisan$^{1}$
and T. V. Ricci$^{2}$
\\
$^{1}$Universidade Federal do Rio Grande do Sul – Departamento de Astronomia – 91501-970, Porto Alegre-RS, Brazil\\
$^{2}$Universidade Federal da Fronteira Sul – Campus Cerro Largo – 97900-000, Cerro Largo-RS, Brazil\\
}

\date{Accepted XXX. Received YYY; in original form ZZZ}

\pubyear{2015}

\begin{document}
\label{firstpage}
\pagerange{\pageref{firstpage}--\pageref{lastpage}}
\maketitle

\begin{abstract}
We investigated the stellar population properties of  a sample of 1\,858 massive compact galaxies (MCGs) extracted from the SDSS survey. Motivated by previous results showing that older compact galaxies tend to have larger velocity dispersion at fixed stellar mass, we used the distance to the $\sigma_\mathrm{e}$ vs. $R_\mathrm{e}$ and $M_\star$ vs. $\sigma_\mathrm{e}$ relations as selection criteria. We found that MCGs are old ($\gtrsim 10$\,Gyr), $\alpha$-enhanced ([$\alpha/\mathrm{Fe}] \sim 0.2$) and have solar to super-solar stellar metallicities. Metallicity increases with $\sigma_\mathrm{e}$, while age and [$\alpha$/Fe] do not vary significantly. Moreover, at fixed $\sigma_\mathrm{e}$, metallicity and stellar mass are correlated. Compared to a control sample of typical quiescent galaxies, MCGs have, on average, lower metallicities than control sample galaxies (CSGs) of similar $\sigma_\mathrm{e}$. For $\sigma_\mathrm{e} \lesssim 225$\,km/s, MCGs are older and more $\alpha$-enhanced than CSGs, while for higher $\sigma_\mathrm{e}$ ages and $\alpha$-enhancement are similar. The differences in age and $\alpha$-enhancement can be explained by lower-$\sigma_\mathrm{e}$ CSGs being an amalgam of quiescent galaxies with a variety of ages. The origin of the differences in metallicity, however, is not clear. Lastly, we compared the stellar mass within the region probed by the SDSS fiber finding that, at fixed fiber velocity dispersion, MCGs have lower stellar masses on average. Since the velocity dispersion is a tracer of the dynamical mass, this raises the possibility that MCGs have, on average, a bottom heavier initial mass function or a larger dark matter fraction within the inner $\sim 1-2$\,kpc.

\end{abstract}




\section{Introduction}

The diversity of galaxies in the local Universe suggests that they follow a variety of evolutionary paths. Despite this diversity, in terms of their star formation activity galaxies can be divided in two main groups: star-forming and quiescent galaxies. Quiescent galaxies are poor in gas or do not convert gas into stars as efficiently as star-forming galaxies. Usually, they are systems with old metal-rich stellar populations that formed in a short period of time (\citealp{Thomas.etal.2005}; \citealp{Graves.etal.2009a}). 

Massive quiescent galaxies ($\log M_\star/M_\odot$\,$\gtrsim$\,10.0) were already common at high redshifts, dominating the high mass end of the stellar mass function by $z \approx 2$ \citep{Davidzon.etal.2017}. However, these high redshift quiescent galaxies are extremely compact compared to their counterparts in the local Universe (\citealp{Damjanov.etal.2009}; \citealp{vanderwel.etal.2014}). Compact quiescent galaxies are extremely rare at $z \sim 0$; their number density dropped by a factor of $\sim$ 20 since $z \sim 1.5$ \citep{Charbonnier.et.al.2017}. One possible explanation for their rarity is that they undergo a two-phase evolutionary process: in the first phase, a compact bulge is formed by wet major mergers or other violent disk instabilities, after which the galaxy rapidly stops its star formation, leaving behind a compact remnant that grows in size by a sequence of dry minor mergers in the second phase. The end result of this process is a giant elliptical galaxy \citep{Naab.etal.2009,Oser.etal.2010,Barro.etal.2013}.  The increase in the fraction of stellar mass stored in the halos of massive early-type galaxies with decreasing redshift \citep{vandokkum10,buitrago17,hill17a} and the presence of metal-poor and $\alpha$-enhanced stars in their outskirts \citep{greene13} support this scenario.

The mean size of star-forming galaxies also increases with time \citep{vanderwel.etal.2014}. This, in turn, contributes to the increase of the mean size of quiescent galaxies, since newly quenched galaxies will tend to be larger. Observations showed that this progenitor bias effect is specially important for $\log M_\star/_\odot \lesssim$\,11.0, where ages and sizes are observed to be anti-correlated \citep{fagioli16,Williams.et.al.2017}. At higher masses there is no clear relation between size and age, pointing to dry minor mergers as the main driver of size evolution \citep{fagioli16,damjanov19}. Consequently, the post-quenching evolution of $z \sim 2$ high mass quiescent galaxies is reasonably well understood, while at lower masses it still is poorly constrained. Growth by dry mergers plays a role \citep{damjanov19}, but another possibility is that some of these galaxies later rejuvenate and acquire a large disk, becoming compact bulges in local spiral and lenticular galaxies \citep{delarosa16,mancini19,constantin22,gao22,hon22}. Alternatively, others will be destroyed in mergers with more massive galaxies \citep{wellons16}, and a few will have a quiet accretion history and evolve passively to the local Universe \citep{Trujillo.etal.2009,yildirim17,Schnorr.et.al.2021}.

How compact quiescent galaxies form and why their star formation was quenched is another topic of intense study. Recent work have shown that at $z \gtrsim 1$ galaxies follow two quenching pathways: one associated with structural changes and short quenching timescales, and another associated with a wide range of quenching timescales and where no significant structural changes occur \citep{wu18,belli19,suess21, tacchella22}. There is evidence that compact quiescent galaxies may have followed either path. Those following the latter path are believed to descend from small star-forming galaxies which experienced little size growth as they increased their stellar mass \citep{vanDokkum.et.al.2015,suess21}, while those following the former likely formed in compaction events \citep{dekel14,zolotov15,lapiner23}, where mergers or violent disk instabilities drive large amounts of gas inwards, triggering a central starburst and reducing the size of the galaxy. 

Despite significant progress, many questions remain unanswered. In particular, little is known about the quenching mechanisms acting in galaxies following each of these paths. AGN feedback and accretion shock heating likely play a role in quenching star formation \citep{dekel06,bluck23}, although their prevalence and relative importance is not clear. As the small sizes and faint stellar continua of $z \sim 2$ quiescent galaxies makes spatially resolved spectroscopic studies of large representative samples prohibitively expensive with current facilities, progress will require a combination of observations of high redshift quiescent galaxies and archaeological studies of passively evolving compact quiescent galaxies at lower redshifts. 

Studies of low redshift ($z \lesssim 0.5$) compact quiescent galaxies have found that they are a mix of young and old objects with super-solar stellar metallicities and [Mg/Fe] \citep{ferre-mateu12,yildirim17,Buitrago.et.al.2018,Spiniello.etal.2021b,Schnorr.et.al.2021}. Compact quiescent galaxies are fast rotators, suggesting that they formed in dissipative events and experienced little to no dry mergers \citep{yildirim17,Schnorr.et.al.2021}. They are found in a variety of environments, however they are more likely to be either central galaxies in the field or satellites in high density environments \citep{Schnorr.et.al.2021,siudek23}, implying that protection from mergers is important for a passive evolution. 

Our understanding of local compact quiescent galaxies is far from complete, however. For example, their photometric and dynamical structures are not yet known in great detail, which hinders our comprehension of their relation to local ellipticals, S0s and compact bulges. Likewise, it is currently not known to which degree young and old compact quiescent galaxies are structurally alike. Recent studies have suggested that old compact galaxies have larger stellar velocity dispersion at fixed stellar mass \citep{Spiniello.etal.2021b,Schnorr.et.al.2021}, which points to significant differences in mass distribution between older and younger compact galaxies. There is tentative evidence that old compact quiescent galaxies have bottom-heavier initial mass function (IMF) slopes \citep{martin-navarro23} than their younger counterparts which could, in theory, account for their larger velocity dispersion. Another possibility is that differences in stellar velocity dispersion are reflecting differences in dark matter halo properties. There is evidence that the stellar velocity dispersion of central quiescent galaxies is proportional to their dark matter halo velocity dispersion \citep{zahid16b,zahid18,sohn20,seo20,utsumi20}, which means that galaxies with higher velocity dispersion at fixed stellar mass occupy more massive or more concentrated dark matter halos \citep{hopkins09}. Lastly, we note that the majority of the studies of low redshift compact quiescent galaxies have been limited to small samples of high mass ($\log M_\star/M_{\odot} \gtrsim 10.8$) extremely compact ($R_\mathrm{e} \lesssim 2$\,kpc) galaxies, whereas the intermediate mass range (10.0 $\lesssim \log M_\star/M_{\odot} \lesssim 10.7$) is still largely unexplored.

In this work, the first in a series of papers, we will characterise the stellar populations properties of a large sample of $z \sim 0$ massive compact quiescent galaxies (MCGs) that are significantly offset from the local $\mathrm{log}\,M_{\star}$ vs. $\mathrm{log}\,\sigma_{\mathrm{e}}$ relation. Our goals are to assess which stellar population properties vary with the distance to the $\mathrm{log}\,M_{\star}$ vs. $\mathrm{log}\,\sigma_{\mathrm{e}}$ relation, and to confirm what has been hinted at by previous studies based on small samples, that compact galaxies which are positive outliers in the $\mathrm{log}\,M_{\star}$ vs. $\mathrm{log}\,\sigma_{\mathrm{e}}$ relation are predominantly old (i.e. mass weighted ages $\gtrsim$ 10\,Gyr). To this end, we will compare age, stellar metallicity and $\alpha$-enhancement of MCGs to a control sample of typical quiescent galaxies, which will also allow us to gain insight into how MCGs are formed and how they relate to the general quiescent galaxy population.

This paper is organised as follows: in Sec.\,\ref{data} we describe the Sloan Digital Sky Survey (SDSS) data used in this paper and our sample selection criteria. In Sec.\,\ref{methods} we present our methods for measuring the stellar population properties. In Sec.\,\ref{sec:results} we present our results and in Sec.\,\ref{sec:discuss} we discuss their implications. Finally, in Sec.\,\ref{sec:conc} we summarise our results and present our conclusions. In this paper we adopt a standard simplified $\Lambda$CDM cosmology  with  $\Omega_{\rm M}$\,=\,0.3, $\Omega_\Lambda$\,=\,0.7 and $H_0$\,=\,70\,km\,s$^{-1}$\,Mpc$^{-1}$.  

\section{Data and Sample Selection}\label{data}

Our sample was extracted from the Sloan Digital Sky Survey (SDSS) data release $14$ \citep{Abolfathi.et.al.2018}. 
Stellar masses ($M_{\star}$) and specific star formation rates (sSFR) used in this work were extracted from the GALEX-SDSS-WISE Legacy Catalog \citep{Salim.etal.2018}. Effective radii ($R_\mathrm{e}$), defined as the semi-major axis of the half-light ellipse, were extracted from the \citet{Simard.etal.2011} catalogue. We used $R_\mathrm{e}$ obtained from Sersic+Exponential fits to the 2D surface brightness profiles of SDSS-DR7 \textit{r}-band images. Stellar velocity dispersions were extracted from the SDSS spectroscopic catalogue. We converted the stellar velocity dispersion (which are measured inside an aperture with a fixed diameter of $3^{\prime \prime}$) to effective velocity dispersion ($\sigma_\mathrm{e}$) using $\sigma_{\mathrm{ap}} = \sigma_\mathrm{e} [R_{\mathrm{ap}}/R_\mathrm{e}]^{-0.066}$ \citep{Cappellari.etal.2006}, where $R_{\mathrm{ap}}$ is the aperture radii. We stress that this correction is small and applying alternative corrections (for example, the one provided by \citealt{zhu23}) does not significantly impact the results presented in this work.

\subsection{Massive Compact Galaxy Sample}

\begin{figure}
\centering
 \includegraphics[width=\columnwidth, trim=10 40 0 10]{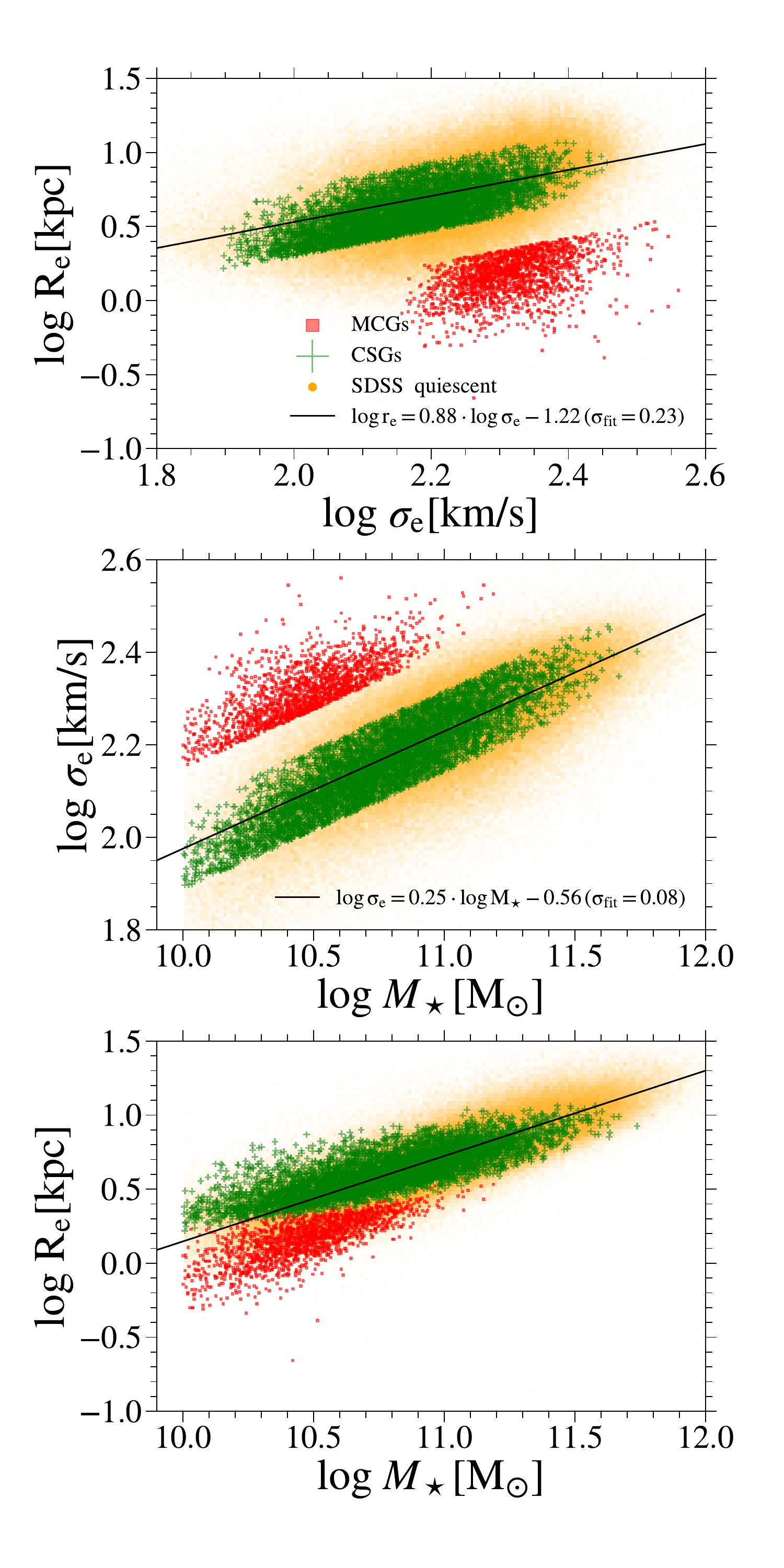}
    \caption{Sample selection.  SDSS quiescent galaxies are showed in orange, MCGs in red, and CSGs in green. Top panel:  $\mathrm{log}\, R_{\mathrm{e}}$ as a function of  $\mathrm{log}\, \sigma_{\mathrm{e}}$ with a linear fit of  $\mathrm{log}\, r_{\mathrm{e}} = 0.88 \cdot  \mathrm{log}\, \sigma_{\mathrm{e}} -1.22$ ($\sigma_{\mathrm{fit}} = 0.23$). Middle panel: $\mathrm{log}\, \sigma_{\mathrm{e}}$ as a function of $\mathrm{log}\, M_{\star}$ and a linear fit of $\mathrm{log}\, \sigma_{\mathrm{e}} = 0.25 \cdot \mathrm{log}\, M_{\star} -0.56$ ($\sigma_{\mathrm{fit}} = 0.08$). Bottom panel:  $\mathrm{log}\, R_{\mathrm{e}}$ as a function of  $\mathrm{log}\, M_{\star}$.}
    \label{fig:final_sample}
\end{figure}

We started by selecting massive quiescent galaxies, defined as those with $\log M_{\star}/M_{\odot} \ge 10$ and $sSFR \le 10^{-11}$ yr$^{-1}$. Next, we performed linear fits to the $\log \sigma_\mathrm{e}$ vs. $\log R_\mathrm{e}$ and $\mathrm{log}\, M_{\star}$ vs. $\mathrm{log}\, \sigma_{\mathrm{e}}$ relations, classifying as massive compact quiescent galaxies (from here on MCGs) those galaxies which lie $2\sigma_{\mathrm{fit}}$ below the fit of the $\mathrm{log}\, \sigma_{\mathrm{e}}$ vs. $\mathrm{log}\, R_{\mathrm{e}}$ relation and $2\sigma_{\mathrm{fit}}$ above the fit of the $\mathrm{log}\, M_{\star}$ vs. $\mathrm{log}\, \sigma_{\mathrm{e}}$ relation. Linear fits were performed with the {\scshape scikit-learn python} package. The results of the fits are shown in Fig.\,\ref{fig:final_sample}. A total of $2\,494$ galaxies were classified as MCGs. Lastly, we restricted our sample to galaxies with $z \le 0.1$, we discarded galaxies with $\sigma_{\mathrm{e}} >380$ km/s (as these high values are either due to poor fits or object superpositions), we removed $9$ galaxies that have bright stars (mag$_\mathrm{r} < 19$) in front of them, and we removed $32$ galaxies that are not included in the \cite{Dominguez.etal.2018} catalogue (although we do not use data from this catalogue in this paper, it will be used in future works with our MCG sample). After these cuts, we are left with a final sample of $1\,858$ MCGs. These are shown as red squares in Fig.\,\ref{fig:final_sample}. Regarding possible biases introduced by our selection criteria, it is worth noting that while measurements of the velocity dispersion in quiescent galaxies have low systematic uncertainties and are accurate even at moderate signal-to-noise \citep{fabricant13,thomas13,zahid16b}, dynamical analyses of fast-rotating early type galaxies have shown that typically $\sigma_\mathrm{z} < \sigma_{R}$ \citep{cappellari16}, meaning that our MCG selection criteria is biased against face-on galaxies. We explored possible effects of this bias and we found that it does not significantly affects our conclusions. For details, see appendix\,\ref{ellipticity_effects}.

Compact galaxies are usually selected based on their size at fixed stellar mass, in contrast to how we select them in this work. To check if galaxies satisfying our selection criteria do indeed have sizes below the local size-mass relation, we plot the distribution of MCGs in the $\mathrm{log}\, M_{\star}$ vs. $\mathrm{log}\, R_{\mathrm{e}}$ diagram, shown at the bottom panel of Fig.\,\ref{fig:final_sample}. MCGs are located below the local relation with very few exceptions, all located around $\log\, M_{\star}/M_{\odot} \lesssim 10.3$, where the size-mass relation of quiescent galaxies flattens \citep{Trevisan.etal.2012,vanderwel.etal.2014}. 

\begin{figure}
\centering
	\includegraphics[width=\columnwidth, trim=10 40 0 10]{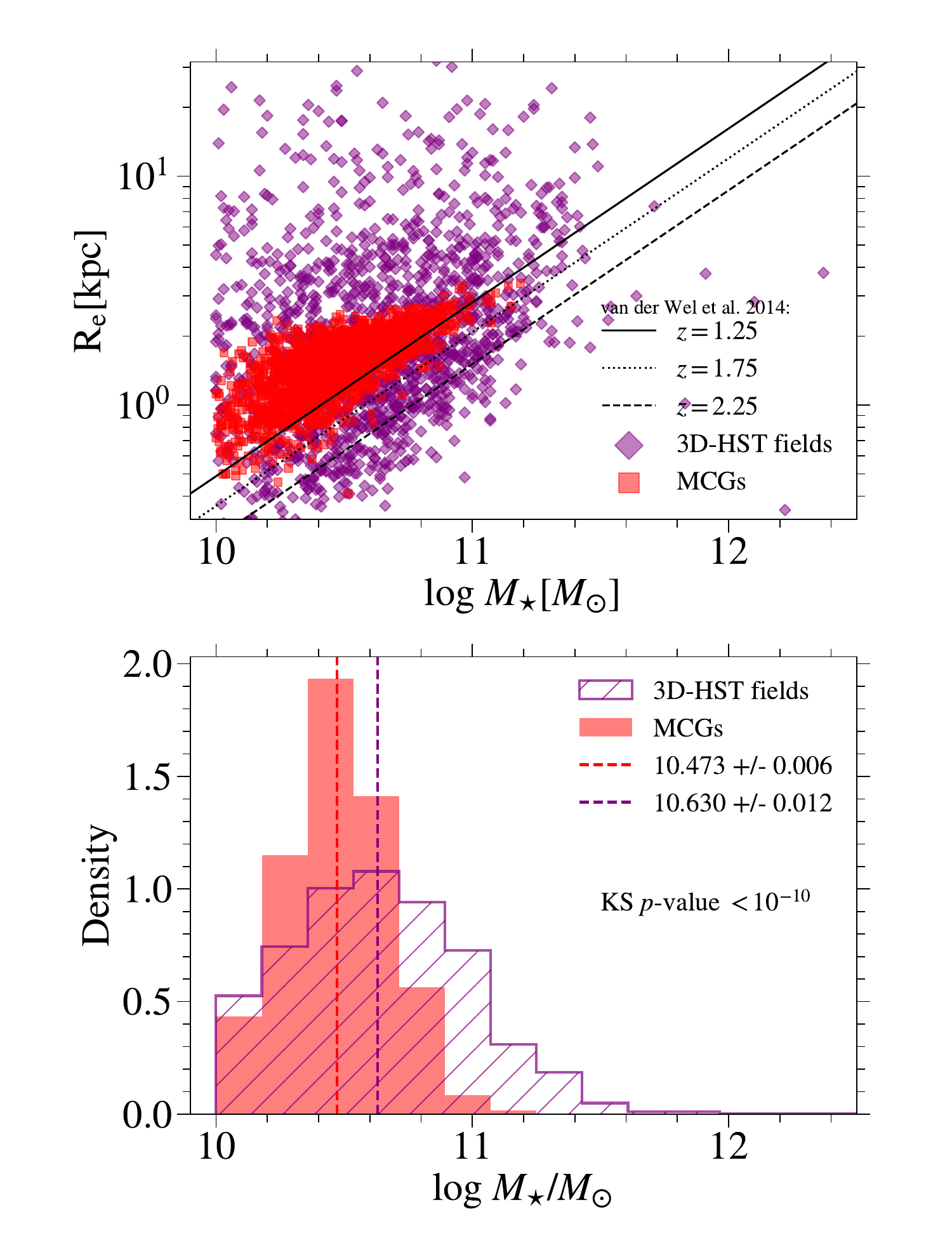}
    \caption{Top panel: log $M_{\star}$ vrs. log $R_\mathrm{e}$ plot for MCGs (red squares) and 3D-HST galaxies (purple diamonds). The solid line ($z = 1.75$), dotted line ($z = 1.75$), and dashed line ($z = 2.25$) are from \protect\cite{vanderwel.etal.2014}. Bottom panel: distribution of stellar mass for MCGs (red) and quiescent galaxies from the 3D-HST fields with redshift between $1.5-3$ (purple). The dashed lines lines indicate the median values of stellar mass for each distribution. }
    \label{fig:MCG_HST}
\end{figure}

Considering that local compact quiescent galaxies are frequently studied as analogues of high redshift ($z \gtrsim 1$) quiescent galaxies, in the top panel of Fig.\,\ref{fig:MCG_HST} we compare the distribution in the $\mathrm{log}\, M_{\star}$ vs. $\mathrm{log}\, R_{\mathrm{e}}$ diagram of MCGs and a sample of $1.5 \leq z \leq 3$ quiescent galaxies extracted from the 3D-HST survey \citep{brammer12,skelton14,momcheva16}. Following \citet{wuyts07} and \citet{williams09}, we classified 3D-HST galaxies as star forming or quiescent based on their rest-frame colours. Galaxy sizes were extracted from \citet{vanderwel.etal.2014}. Overall, MCGs overlap with the distribution of high redshift quiescent galaxies, although MCGs are clearly shifted to larger sizes at fixed stellar mass, with few lying close to the $z = 1.75$ size-mass relation. In the bottom panel of Fig.\,\ref{fig:MCG_HST} we compare the mass distribution of the samples. It is noticeable that, compared to the distribution of high redshift quiescent galaxies, the MCG distribution is shifted to lower stellar masses and that the fraction of high mass MCGs ($\mathrm{log}\, M_{\star}/M_\odot \gtrsim 10.8$) is much smaller. Nonetheless, quiescent galaxies with similar sizes to MCGs are present in all redshift bins,  as shown in Fig. $5$ from \citet{vanderwel.etal.2014}, so a connection between MCGs and high redshift quiescent galaxies cannot be discarded based on their sizes alone.

Lastly, due to the differences in the size-mass distribution of MCGs and high redshift quiescent galaxies, we decided to test if adopting an additional selection criterion based on the distance to the size-mass relation might affect our results. We found that it does not have a significant impact on our conclusions (see appendix\,\ref{sec:mass-size}). This is not surprising, as previously it has been shown that, in local early type galaxies, no stellar population property (measured from SDSS spectra) shows any dependence on $R_\mathrm{e}$ at fixed velocity dispersion \citep{Graves.etal.2009}. Understanding the reason for the large variation in size among MCGs with similar stellar population properties is beyond the scope of this paper and we leave it for future work.

\begin{figure}
\centering
	\includegraphics[width=0.9\columnwidth, trim=10 40 0 10]{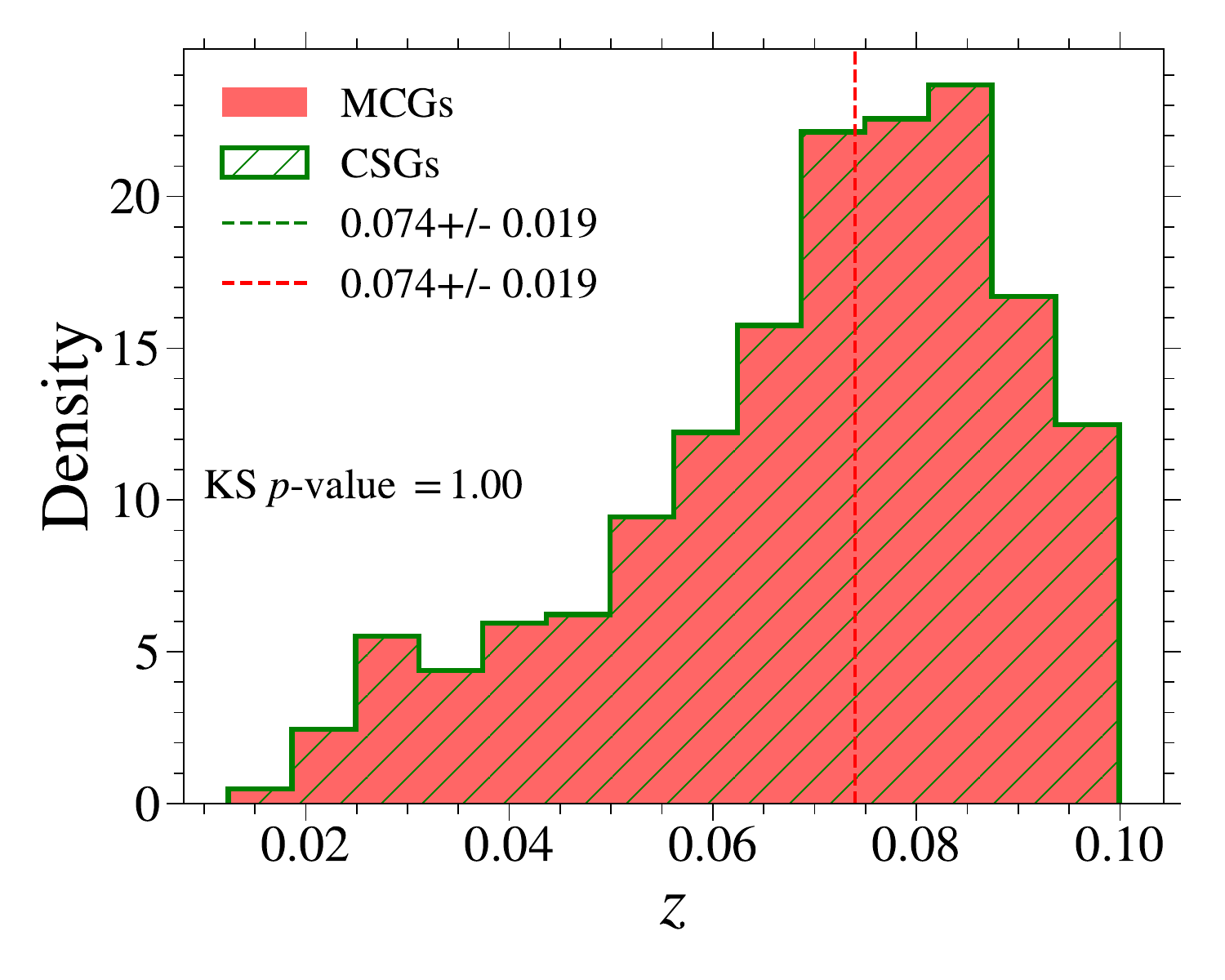}
    \caption{Comparison between the redshift of MCGs (red filled histogram)  and CSGs (green step histogram). The dashed lines indicate the median values of each sample.}
    \label{fig:pvalue_sample}
\end{figure}

\begin{figure}
\centering
	\includegraphics[width=\columnwidth, trim=10 40 0 10]{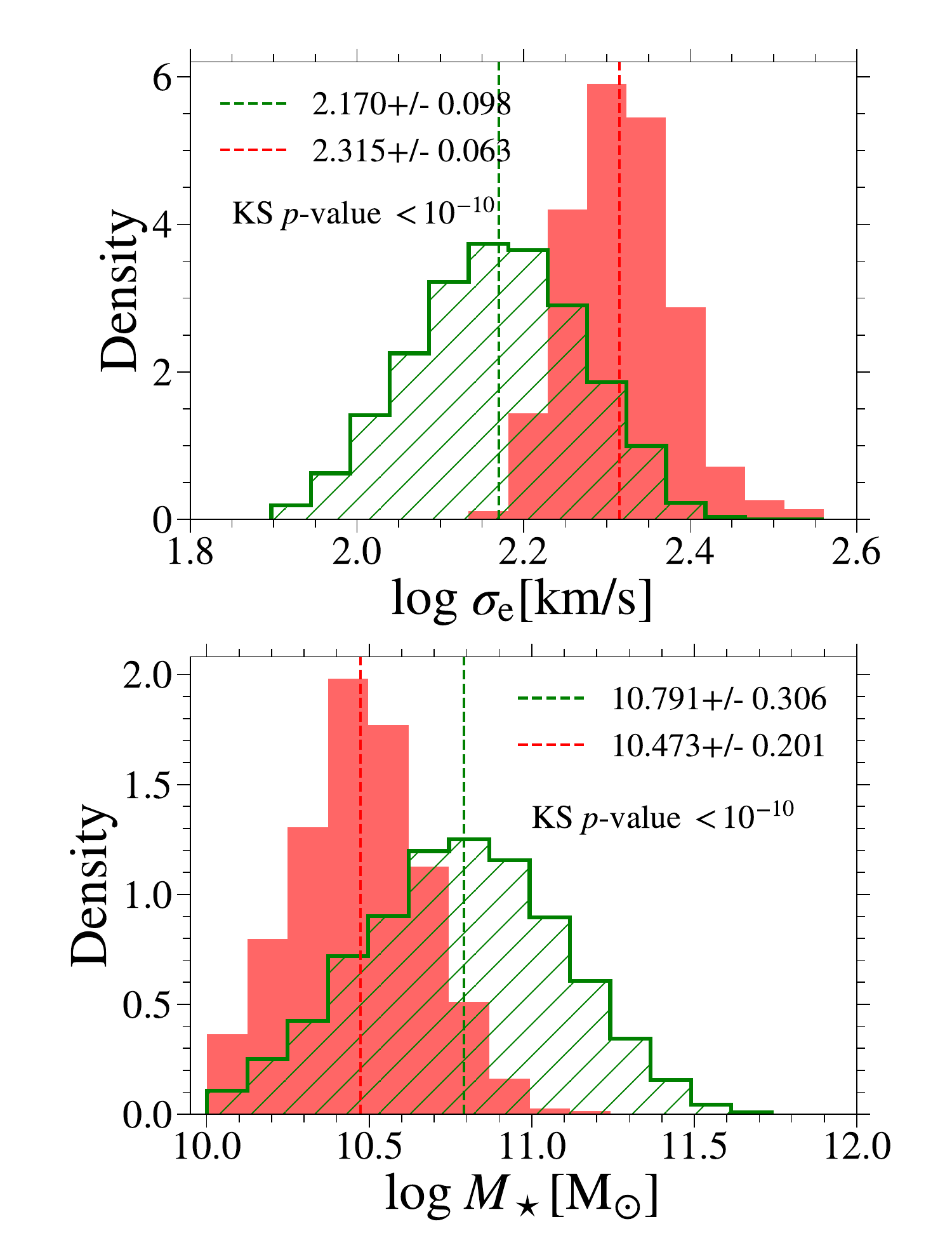}
    \caption{Top panel: $\sigma_\mathrm{e}$ distribution of MCGs (in red) and CSGs (in green). Bottom panel: $M_{\star}$ distribution. The dashed lines are the median values of each distribution.}
    \label{fig:stellar_mass_sigma}
\end{figure}

\subsection{Control Sample}
\label{sec:normal_quiescent_sample}

In order to understand how the star formation history of MCGs differs from that of typical quiescent galaxies, we  compared their stellar population properties with those of a control sample of quiescent galaxies that lies within $\pm1\sigma_{\mathrm{fit}}$ of the linear fits to the $\log \sigma_\mathrm{e}$ vs. $\log R_\mathrm{e}$ and $\mathrm{log}\, M_{\star}$ vs. $\mathrm{log}\, \sigma_{\mathrm{e}}$ relations. The samples were matched in redshift to account for the fixed fibre diameter of $3^{\prime \prime}$. By matching the samples in redshift we guarantee that the SDSS spectra sample a region of the same size (the inner $\sim 1-3$\,kpc) in galaxies in both samples. This is important because our goal is to probe the stellar populations formed in situ which are predominantly located in the inner few kpc, while the contribution of accreted stars to the light budget increases with radius. The matching procedure was performed with the {\scshape MatchIt R} package \citep{Ho.et.al.2011} using the Propensity Score Matching (PSM) technique \citep{Rosenbaum.and.Rubin.1983,DeSouza.etal.2016}. The control sample has three times the size of our MCG sample, totalling $5\,574$ galaxies. Control sample galaxies (CSGs) are shown as green crosses in Fig.\,\ref{fig:final_sample}. 

\begin{figure}
\centering
	\includegraphics[width=0.9\columnwidth, trim=10 40 0 10]{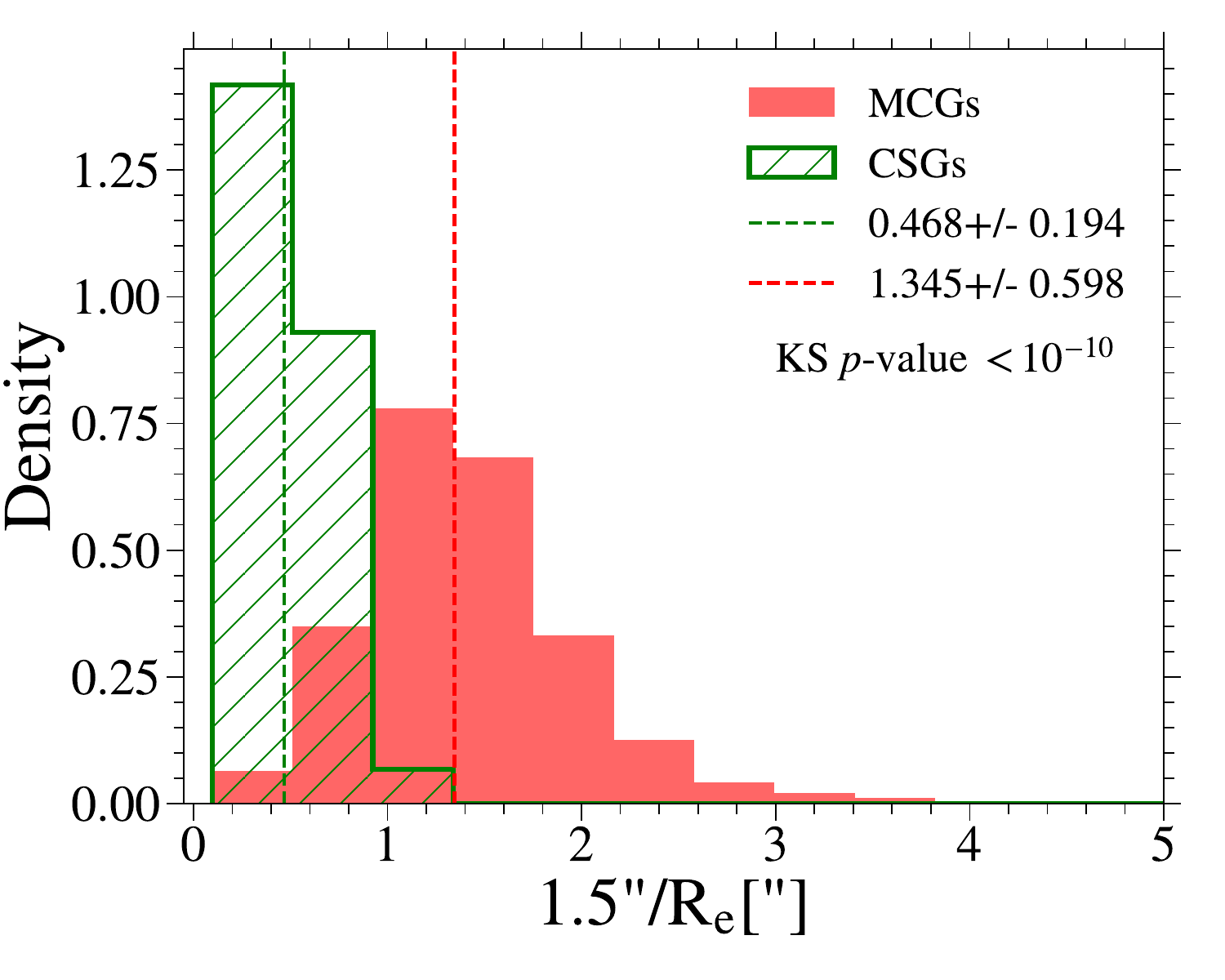}
    \caption{Comparison between the distribution of fiber angular radii (in units of $R_\mathrm{e}$) probed by the SDSS spectra.}
    \label{fig:fibra}
\end{figure}

\begin{figure*}
\centering
	\includegraphics[width=\textwidth]{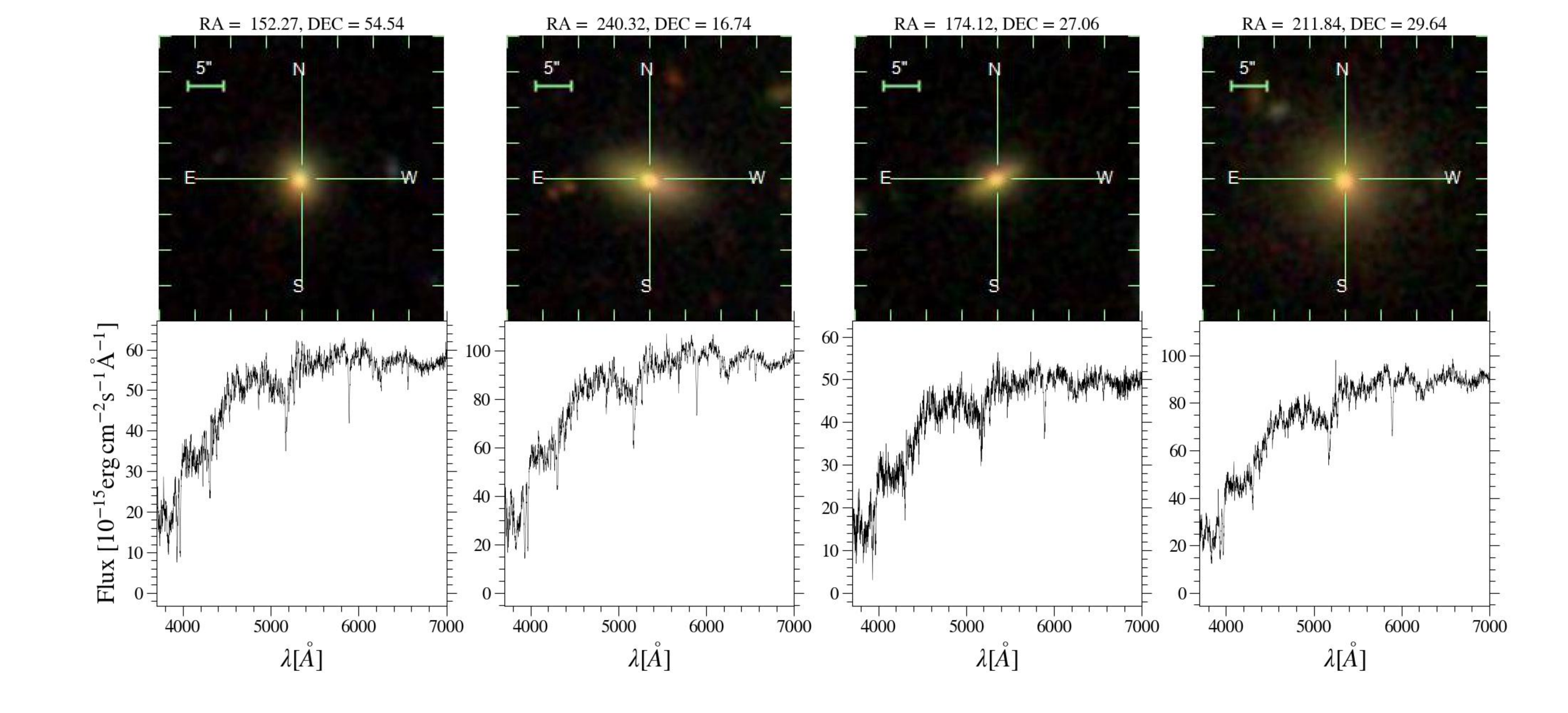}
    \caption{SDSS images and spectra of four MCGs. Images are shown in the top panels and the respective spectrum of each galaxy in the bottom panel. RA $= 152.27$, DEC $= 54.54$: $R_\mathrm{e} = 0.92$ kpc, $\mathrm{log}\, M_{\star}/M_{\odot} = 10.37$, and $\sigma_{\mathrm{e}} = 178.29$ km/s; RA $= 240.32$, DEC $= 16.74$: $R_\mathrm{e} = 1.29$ kpc, $\mathrm{log}\, M_{\star}/M_{\odot} = 10.46$, and $\sigma_{\mathrm{e}} = 214.74$ km/s;  RA $= 174.11$, DEC $= 27.06$: $R_\mathrm{e} = 1.95$ kpc, $\mathrm{log}\, M_{\star}/M_{\odot} = 10.71$, and $\sigma_{\mathrm{e}} = 263.57$ km/s; and RA $= 211.84$, DEC $= 29.64$: $R_\mathrm{e} = 2.74$ kpc, $\mathrm{log}\, M_{\star}/M_{\odot} = 10.99$, and $\sigma_{\mathrm{e}} = 300.41$ km/s.}
    \label{fig:gal_spectra_MCGs}
\end{figure*}

\begin{figure*}
\centering
	\includegraphics[width=\textwidth]{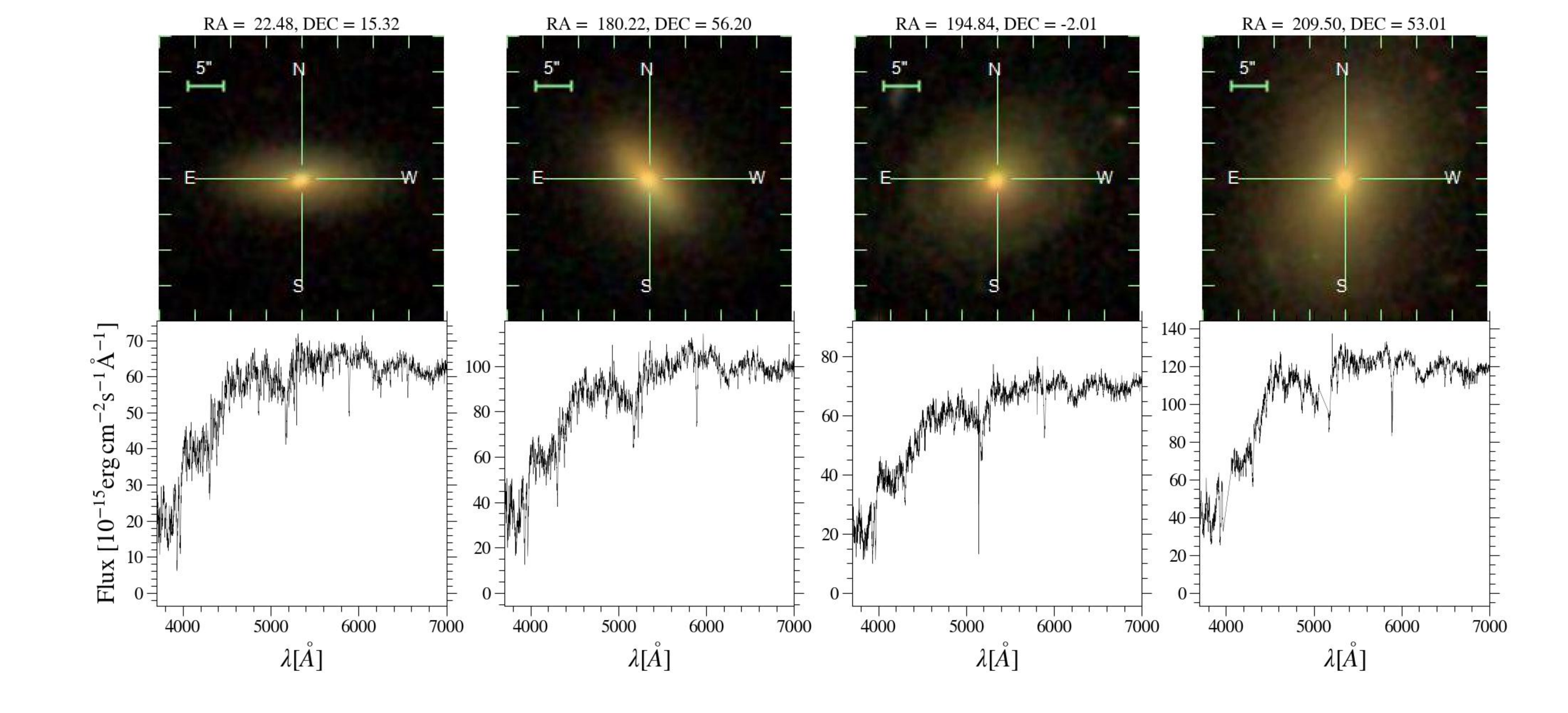}
    \caption{SDSS images and spectra of four CSGs, similar to Fig. \ref{fig:gal_spectra_MCGs}. RA $= 22.48$, DEC $= 15.32$: $R_\mathrm{e} = 3.87$ kpc, $\mathrm{log}\, M_{\star}/M_{\odot} = 10.77$, and $\sigma_{\mathrm{e}} = 140.51$ km/s;  RA $= 180.22$, DEC $= 56.20$: $R_\mathrm{e} = 3.85$ kpc, $\mathrm{log}\, M_{\star}/M_{\odot} = 11.21$, and $\sigma_{\mathrm{e}} = 193.94$ km/s;  RA $= 194.84$, DEC $= -2.01$: $R_\mathrm{e} = 7.0$ kpc, $\mathrm{log}\, M_{\star}/M_{\odot} = 11.38$, and $\sigma_{\mathrm{e}} = 248.64$ km/s; and  RA $= 209.50$, DEC $= 53.01$: $R_\mathrm{e} = 8.6$ kpc, $\mathrm{log}\, M_{\star}/M_{\odot} = 11.62$, and $\sigma_{\mathrm{e}} = 285.94$ km/s.}

    \label{fig:gal_spectra_CSGs}
\end{figure*}

In Fig.\,\ref{fig:pvalue_sample} we show the redshift distribution of the two samples. In Fig.\,\ref{fig:stellar_mass_sigma} we show a comparison of the $\sigma_\mathrm{e}$ and $M_{\star}$ distributions of MCGs and CSGs. We note that, due to our selection criteria, MCGs have higher $\sigma_\mathrm{e}$ and lower $M_{\star}$ than CSGs. In Fig.\,\ref{fig:fibra}, we compare the radii (in units of $R_\mathrm{e}$) probed by the SDSS spectra. The spectra of CSGs probe a region smaller than $1\,R_\mathrm{e}$ for the majority of galaxies, while the spectra of MCGs probe $\sim 1-2\,R_\mathrm{e}$. This difference needs to be taken into account when interpreting our results. We further discuss this in Sec.\,\ref{sec:accretion}. In Fig.\,\ref{fig:gal_spectra_MCGs} and \ref{fig:gal_spectra_CSGs} we show examples of typical MCGs and CSGs respectively.

We note that we did not apply any cuts to the MCGs and CSGs samples in order for them to be complete down to the r-band absolute magnitude ($M_r$) limit. We investigated the effects of incompleteness by applying a cut in $M_r < -20.4$, concluding that our results do not change significantly. This is further discussed in appendix \ref{sec:completeness_sdss}.

\section{Methodology}\label{methods}

\begin{figure}
\centering
	\includegraphics[width=\columnwidth, trim=10 40 0 10]{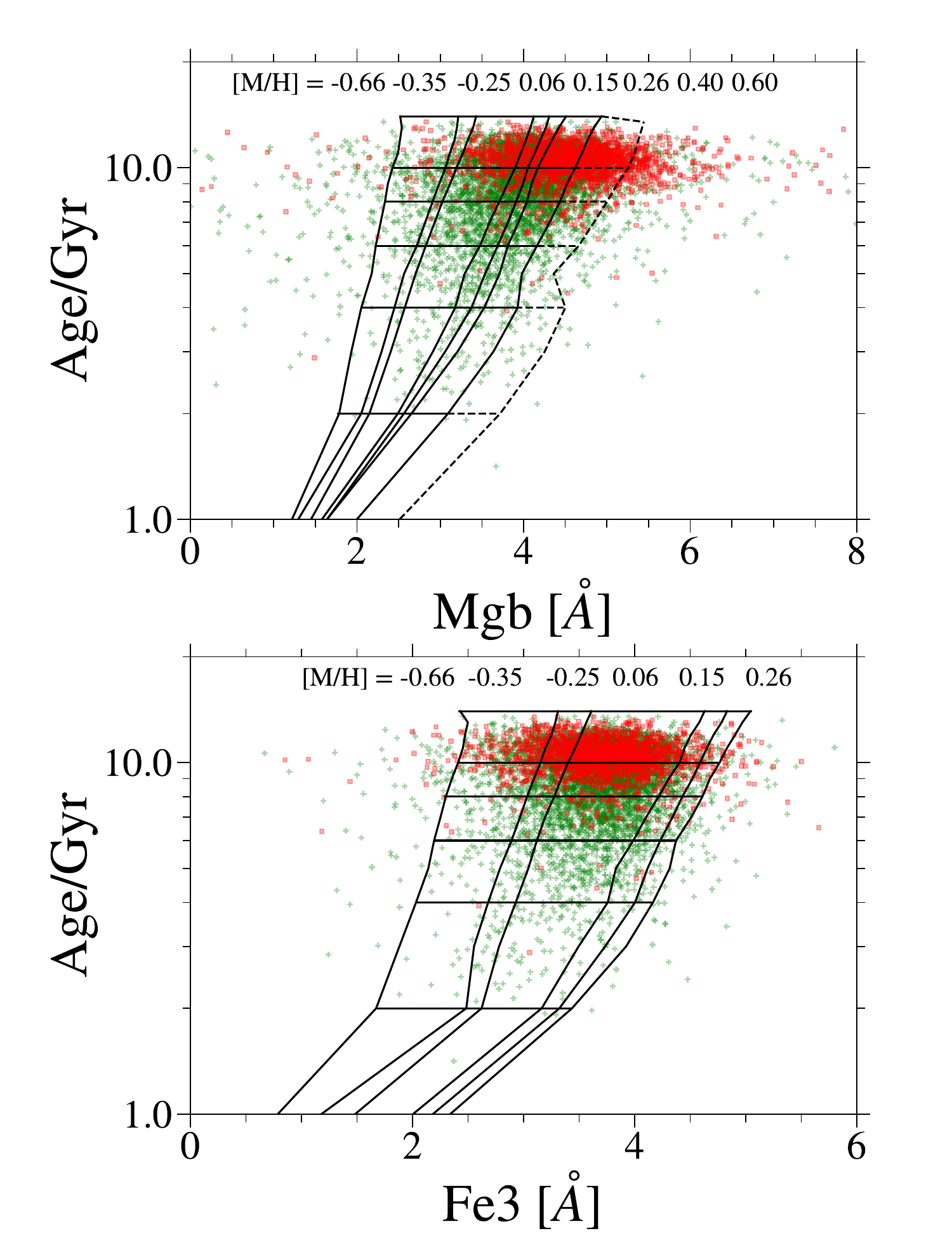}
    \caption{Top panel:Luminosity-weighted age as a function of Mgb. The solid lines are the V15 models and the dashed line is the metallicity extrapolated to [M/H] = $+0.6$. Bottom panel: Luminosity-weighted age as a function of spectral index Fe3.}
    \label{fig:metodology_alpha_fe}
\end{figure}

 The comparison of stellar population properties of MCGs and CSGs, consists of two approaches: a comparison of spectral indices and of properties derived from stellar population synthesis.

\subsection{Spectral indices}
\label{spectral_indices}

Spectral indices are continuum regions or stellar absorption features of the spectra that represent direct observations of the stellar population properties of galaxies. $25$ strong absorption features are summarized in the Lick system \citep{Worthey1994} with well defined central ‘feature bandpass’ and two ‘pseudo-continuum bandpasses’ in a  resolution of $\approx 8$ \AA. In addition, there are spectral indices such as D$_\mathrm{n}4000$ \citep{Balogh.et.al.1999}, which measures the strength of the discontinuity in the spectrum around $4000$\,\AA, and H$\alpha_A$ \citep{Nelan.et.al.2005}, which are not part of the Lick system. 

Each spectral index contains information about age, metallicity, or [$\alpha$/Fe], and frequently the same index might provide information about more than one of these properties. For example, Dn$4000$ is sensitive to both age and metallicity, although its dependence on age is stronger \citep{Kauffmann.etal.2003b}. Other indices sensitive to age are those that measure the strength of Balmer absorption lines such  as H$\delta_A$, H$\gamma_A$, H$\beta$ and H$\alpha$, although H$\delta_A$ and H$\gamma_A$ also show a significant dependence on [$\alpha$/Fe], as reported by \cite{Thomas.et.al.2003b}. 

Lick indices that measure the strength of prominent metal absorption lines show a dependence on both metallicity and  [$\alpha$/Fe]. To investigate the metallicity separately from [$\alpha$/Fe], \cite{Bruzual.and.Charlot.2003} and \cite{Thomas.etal.2003} defined the combined indices [Mg$2$Fe] $= 0.6 \mathrm{Mg}_2 + 0.4\cdot	 \log(\mathrm{Fe}4531 + \mathrm{Fe}5015)$ and [MgFe]$^\prime$ $= \sqrt{\mathrm{Mgb} \cdot	 (0.72 \cdot	 \mathrm{Fe}5270 + 0.28 \cdot	\mathrm{Fe}5335)}$  respectively. To estimate [$\alpha$/Fe], the ratio Mgb/$\langle$Fe$\rangle$, with $\langle$Fe$\rangle$ = (Fe5270+Fe5335)/2,  is typically employed \citep{Thomas.etal.2003}. 

In this paper, we measured the Dn$4000$, H$\beta$, H$\alpha$, Mgb/Fe3\footnote{Fe$3$ = (Fe4383 + Fe5270 + Fe5335)/3 \citep{Kuntschner.et.al.2001}}, [Mg$2$Fe], and [MgFe]$^\prime$ spectral indices to assess differences between the stellar population properties of MCGs and CSGs. Spectral indices were measured by running the publicly available code {\scshape pylick} \citep{Borghi.etal.2022} on emission line subtracted spectra (see Sec.\,\ref{sec:spectral_synthesis} for details on the emission line subtraction). As spectral indices are sensitive to the internal velocity dispersion of the galaxy \citep{Trager.etal.1998,delaRosa.etal.2007}, we derive a broadening correction function to account for this effect. We followed a similar methodology  to \citet{delaRosa.etal.2007}. More details on this correction can be found in appendix \ref{sec:broadening_function}. Uncertainties on spectral indices values were determined by running {\scshape pylick} on $1\,000$ realisations of the observed spectrum. We adopt as uncertainties the standard deviation of the spectral indices values. 

\subsection{Spectral synthesis and emission line subtraction}
\label{sec:spectral_synthesis} 

To estimate the age and stellar metallicity of galaxies we used the penalized pixel fitting code {\scshape ppxf} \citep{Cappellari.and.Eric.2004}. Although {\scshape ppxf} can fit emission and absorption lines simultaneously, considering that emission lines in MCGs are faint and their accurate subtraction is crucial for the measurement of spectral indices, we opt to perform the emission line subtraction ourselves. To this end, we first ran {\scshape ppxf} with the goal of subtracting the stellar continuum to identify and fit any emission lines that were detected. 

We fitted the observed stellar continuum in the  $3900-7400$ \AA ~ range (normalised in $5635$ \AA) using the \cite[V15]{Vazdekis.etal.2015} simple stellar population (SSP) models as templates. The V15 models were computed with a \cite{Kroupa.etal.2001} initial mass function (IMF) and stellar evolution tracks from Bag of Stellar Tracks and Isochrones (BaSTI, \citealt{Pietrinferni.etal.2004,Pietrinferni.etal.2006}). We used as templates models with $15$ different ages, varying between $30$ Myr and $13.5$ Gyr, $6$ different metallicity values, varying between [M/H] = $-1.3$ and $+0.4$, and [$\alpha$/Fe] $=$ BaseFe (BaseFe models are those computed under the assumption that the MILES stars have solar $\alpha$-abundances). We ran {\scshape ppxf} using four Gauss-Hermite moments and with no additive or multiplicative Legendre polynomials. The spectral regions around prominent emission lines such as [O\,III], [N\,II], [S\,II] and the Balmer lines were masked out. The resulting synthetic spectra was subtracted from the observed spectra and the residuals were inspected for the presence of emission lines.

We fitted the H$\delta$, H$\gamma$, H$\beta$, H$\alpha$, [O {\scshape iii}]$\lambda 5007$, and [N{\scshape ii}]$\lambda 6584$ emission lines using the publicly available {\scshape ifscube} code \citep{Ruschel-Dutraetal.2021}. We fitted the continuum in the range of $4040-7200$~\AA~ with a 10th degree polynomial and we fitted the emission lines with Gaussian profiles. Two Gaussian components were used to fit H$\alpha +$ [N {\scshape ii}] and one Gaussian component was used for the other emission lines.

Uncertainties in the measurement of emission lines were estimated from Monte Carlo (MC) simulations. In MC we used a normal distribution with $100$ realisations of the residual spectrum (observed - synthetic spectrum), where $\sigma_{\mathrm{err}}$ is the error spectrum. For each realisation, emission-lines fluxes were measured with  {\scshape ifscube}. Emission line uncertainties were defined as the standard deviation of the emission-line
fluxes from the $100$ realisations. We calculated the signal-to-noise ratio (SNR) by dividing the nominal flux by the uncertainty. We used SNR $\geq 3$ as a criteria to select reliable emission lines fits which were in turn subtracted from the observed spectra.
Next, we ran {\scshape ppxf} on the emission line subtracted spectra, this time fitting the stellar continuum in the 3900 to 5800\AA~ range to avoid IMF-sensitive features. Mass-weighted ages and metallicities of the best fitting model were calculated using the following equations: 

\begin{equation}
    \mathrm{log} \, t_{\mathrm{MW}} = \frac{\sum_{j = 1}^{n} \, \mu_j \, \mathrm{log} \, t_j }{\sum_{j = 1}^{n} \, \mu_j}
\end{equation}

\begin{equation}
    \mathrm{[M/H]}_{\mathrm{MW}} = \frac{\sum_{j = 1}^{n} \, \mu_j \, \mathrm{[M/H]}_j}{\sum_{j = 1}^{n} \, \mu_j}
\end{equation}

where $n$ is the number of models, $\mu_j$, $t_j$, and  $\mathrm{[M/H]}_j$, are the 
fraction contribution of total stellar mass, age, and metallicity of each SSP model, respectively. We also added the lookback time of each galaxy to the derived ages.

\subsubsection{[$\alpha$/Fe]}

To estimate [$\alpha$/Fe], we followed the approach described in \cite{LaBarbera.et.al.2013} and \cite{Vazdekis.etal.2015},  which is based on the spectral indices Mg$b$ and Fe3. For each galaxy, we estimate two independent metallicities, $Z_{{\rm Mg}b}$ and $Z_{{\rm Fe}3}$, by fixing the galaxy age and interpolating the model grid. This procedure is illustrated in Fig.~\ref{fig:metodology_alpha_fe}, where we show the galaxy luminosity-weighted ages as a function of Mg$b$ and ${\rm Fe}3$, as well as the predictions from the V15 models with different metallicities. As discussed by \cite{LaBarbera.et.al.2013}, estimating $Z_{{\rm Mg}b}$ of an $\alpha$-enhanced population may require extrapolation of the models to higher metallicities. This is illustrated in the upper panel of Fig.~\ref{fig:metodology_alpha_fe}, where we show our linear extrapolation of the model Mg$b$ to metallicity $[{\rm Z}/{\rm H}] = +0.6$. The proxy of [$\alpha$/Fe] is then defined as the difference between these metallicities, $[Z_{{\rm Mg}b} / Z_{{\rm Fe}3}] = Z_{{\rm Mg}b} - Z_{{\rm Fe}3}$, and to obtain $[\alpha/{\rm Fe}]$ we use the relation \citep{Trevisan.et.al.2017}:
\begin{equation}
     [\alpha/{\rm Fe}] = -0.07 + 0.51\ [Z_{{\rm Mg}b} / Z_{{\rm Fe}3}]
\end{equation} 

\section{Results}
\label{sec:results}

\begin{figure*}
\centering
	\includegraphics[width=\textwidth]{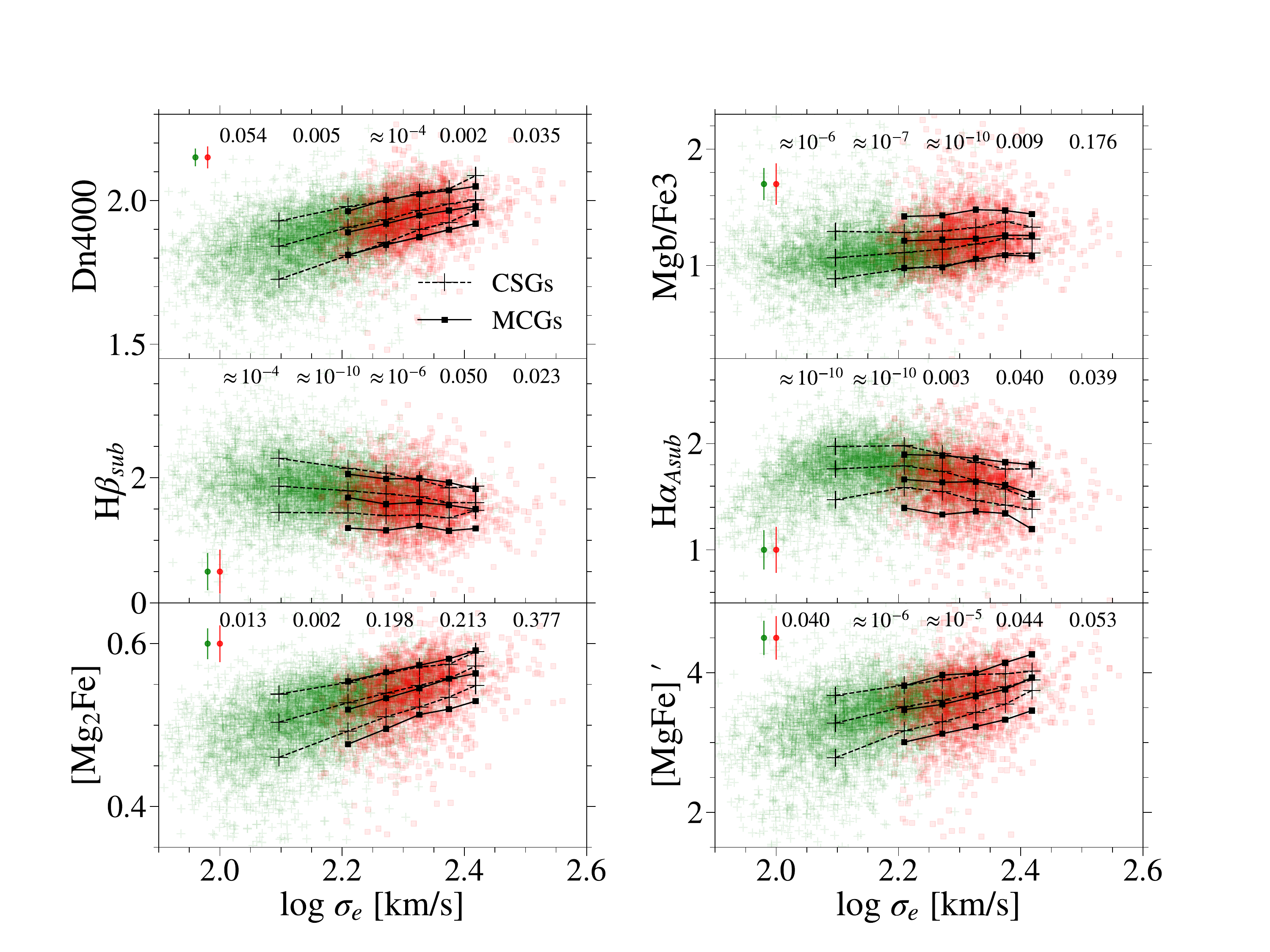}
    \caption{Spectral indices as a function of $\sigma_\mathrm{e}$. MCGs are shown in red and CSGs in green. The $16$th, $50$th, and $80$th percentiles of the MCG distribution in each bin of $\sigma_\mathrm{e}$ are shown as square markers connected by solid lines, while CSGs percentiles are shown as cross markers connected by dashed curves. The values at the top of each panel are the KS $p-$value for each bin (significance level: $5\%$).}
    \label{fig:indices_pylick}
\end{figure*}

\subsection{Spectral indices}

 Previous works have shown that the stellar population properties of early-type galaxies correlate better with velocity dispersion than with other parameters such as stellar mass or luminosity (\citealp{Graves.etal.2009a,Graves.etal.2009}; \citealp{vanderwel.et.al.2009}), indicating  that  velocity dispersion is a better predictor of the galaxies' star formation histories (SFHs). Thus, in Fig.\,\ref{fig:indices_pylick} we show the values of the spectral indices D$_n$4000, H$\beta$, H$\alpha_A$, [Mg$_2$Fe], [MgFe]$^\prime$, and Mg$_b$/Fe3 as a function of $\sigma_\mathrm{e}$. The  samples were divided into six bins of $\sigma_\mathrm{e}$ for which we calculated the $16$th, $50$th, and $84$th percentiles of the distributions. Galaxies with $150\,\mathrm{km/s} < \sigma_\mathrm{e} < 250$\,km/s  were divided into four bins with a width of $25$\,km/s. The remaining two bins contain galaxies with  $\sigma_\mathrm{e} < 150$\,km/s and $\sigma_\mathrm{e} > 250$\,km/s. 

The age sensitive indices H$\beta$ and H$\alpha$ show similar trends. In the two lowest $\sigma_\mathrm{e}$ bins MCGs have lower median values (older ages) than CSGs, while in the remaining bins the medians are similar. In all bins MCGs have lower $16$th percentile values, although the differences in the two highest $\sigma_\mathrm{e}$ bin are marginal. The metallicity sensitive indices [Mg$_2$Fe] and [MgFe]$^\prime$  show a similar behaviour in which the median and 84th percentile of MCGs and CSGs are similar, but in MCGs the 16th percentile is shifted to smaller values (lower metallicities) in all bins. Regarding the [$\alpha$/Fe] sensitive index ratio Mgb/Fe$3$, MCGs have larger median and 84th percentile values in all bins, although the difference is not statistically significant in the $\sigma_\mathrm{e} > 250$\,km/s bin. In contrast, 16th percentile values are similar. The trends in D$_n4000$ are harder to interpret, as the value of this index increases with both increasing age and metallicity and, to a lesser degree, decreases with increasing [$\alpha$/Fe]. Nonetheless, considering the trends in the other indices discussed above, MCGs displaying lower median values in all bins is likely due to a combination of lower metallicities and higher [$\alpha$/Fe] in MCGs.

In summary, based on a comparison of spectral indices, we conclude that MCGs tend to be older, have lower metallicities and higher $\alpha$-enhancement than CSGs, although differences in [$\alpha$/Fe] are not statiscally significant for $\sigma_\mathrm{e} \gtrsim 250$\,km/s. We note that the difference between the 16th percentile of the distributions (or 84th percentile in the case of Mgb/Fe$3$) are largest than the differences between the medians.

\subsection{Spectral synthesis}\label{synthesis}

\begin{figure}
\centering
	\includegraphics[width=\columnwidth, trim=10 40 0 10]{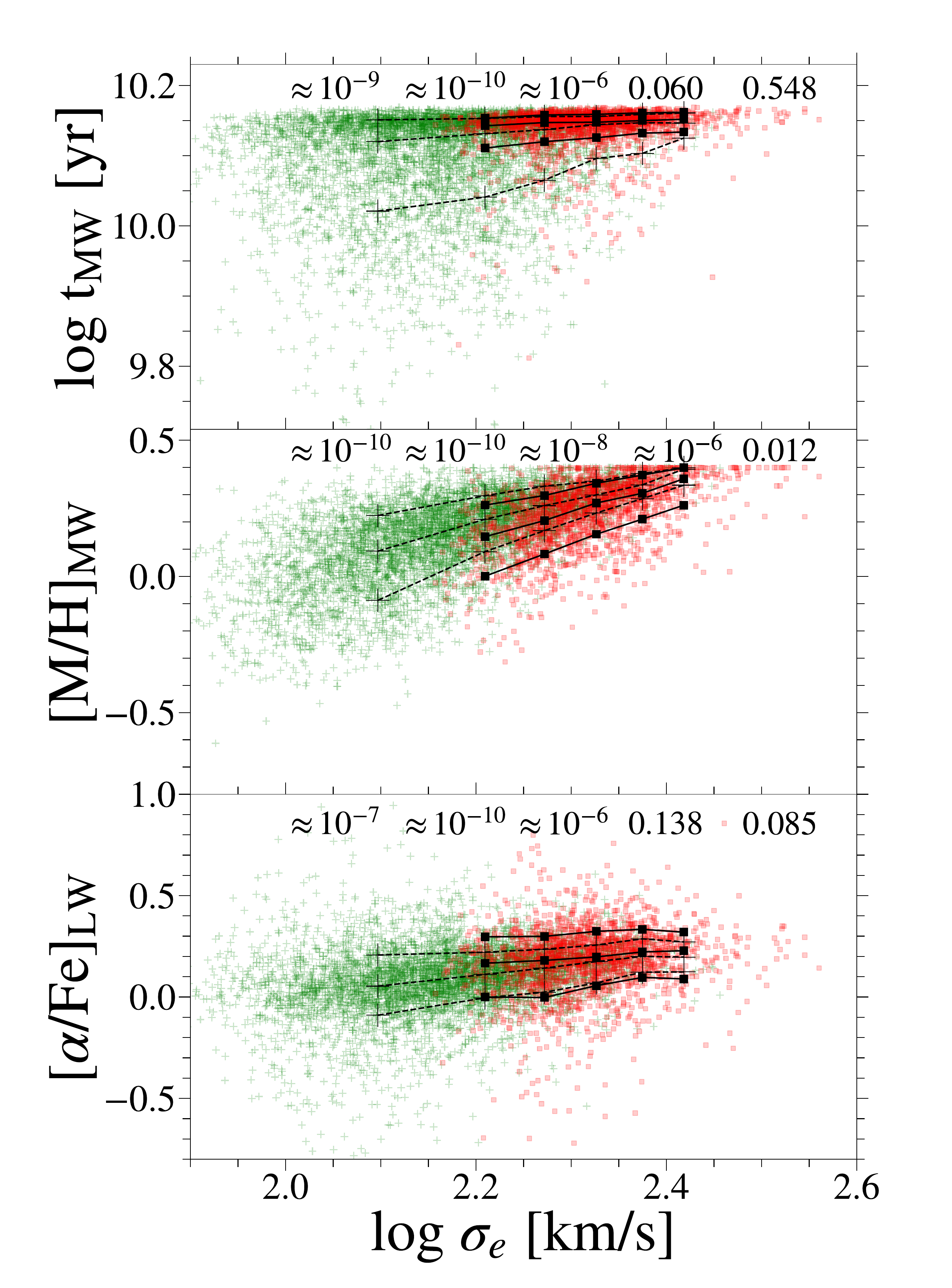}
    \caption{Stellar population properties as a function of $\sigma_\mathrm{e}$ for MCGs (in red) and CSGs (in green). Top panel: mass-weighted ages. Middle panel: mass-weighted metallicities. Bottom panel: luminosity-weighted [$\alpha$/Fe]. The percentile, bins, and lines are the same as in Fig.\,\ref{fig:indices_pylick}. The values at the top of each panel are the KS $p-$value tests results for each bin. Note that, we added the lookback time to estimated the final mass-weighted ages.}
    \label{fig:age_metallicity_alpha_fe_ppxf}
\end{figure}

\begin{figure}
\centering
	\includegraphics[width=\columnwidth, trim=10 40 0 10]
 {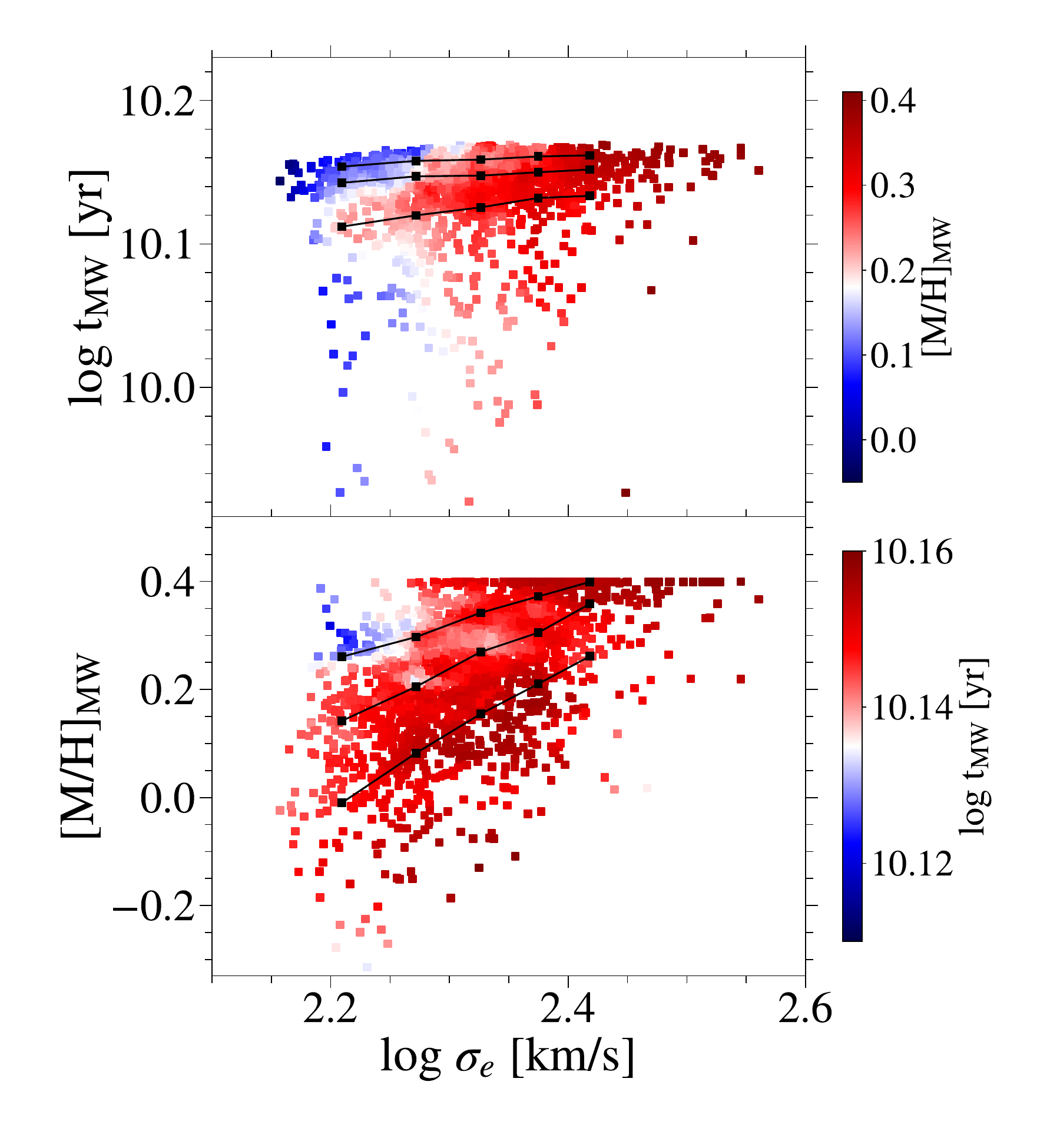}
    \caption{Mass-weighted age and metallicity of MCGs as a function of $\sigma_\mathrm{e}$ coloured by the mass-weighted age and  metallicity. Each point was coloured according to its LOESS smoothed value.}
    \label{fig:stellar_properties_loess}
\end{figure}

\begin{figure*}
\centering
	\includegraphics[width=\textwidth]{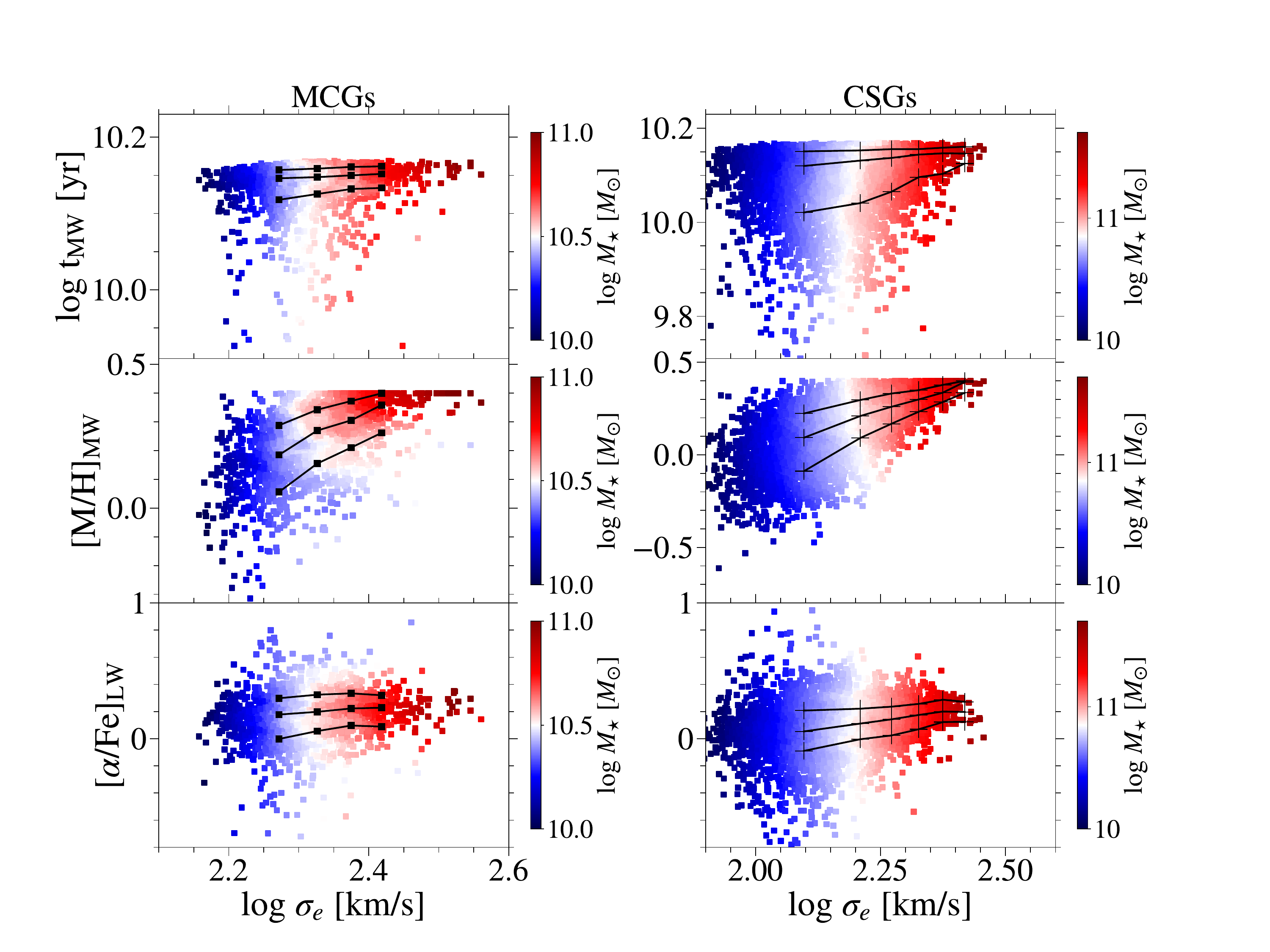}
    \caption{Mass-weighted age, metallicity, and luminosity-weighted [$\alpha$/Fe] of MCGs (left panels) and of CSGs (right panels) as a function of $\sigma_\mathrm{e}$. Each point was coloured according to its LOESS smoothed $M_{\star}$.}
    \label{fig:stellar_properties_loess_stellar_mass}
\end{figure*}

\begin{figure}
\centering
	\includegraphics[width=\columnwidth, trim=10 40 0 10]{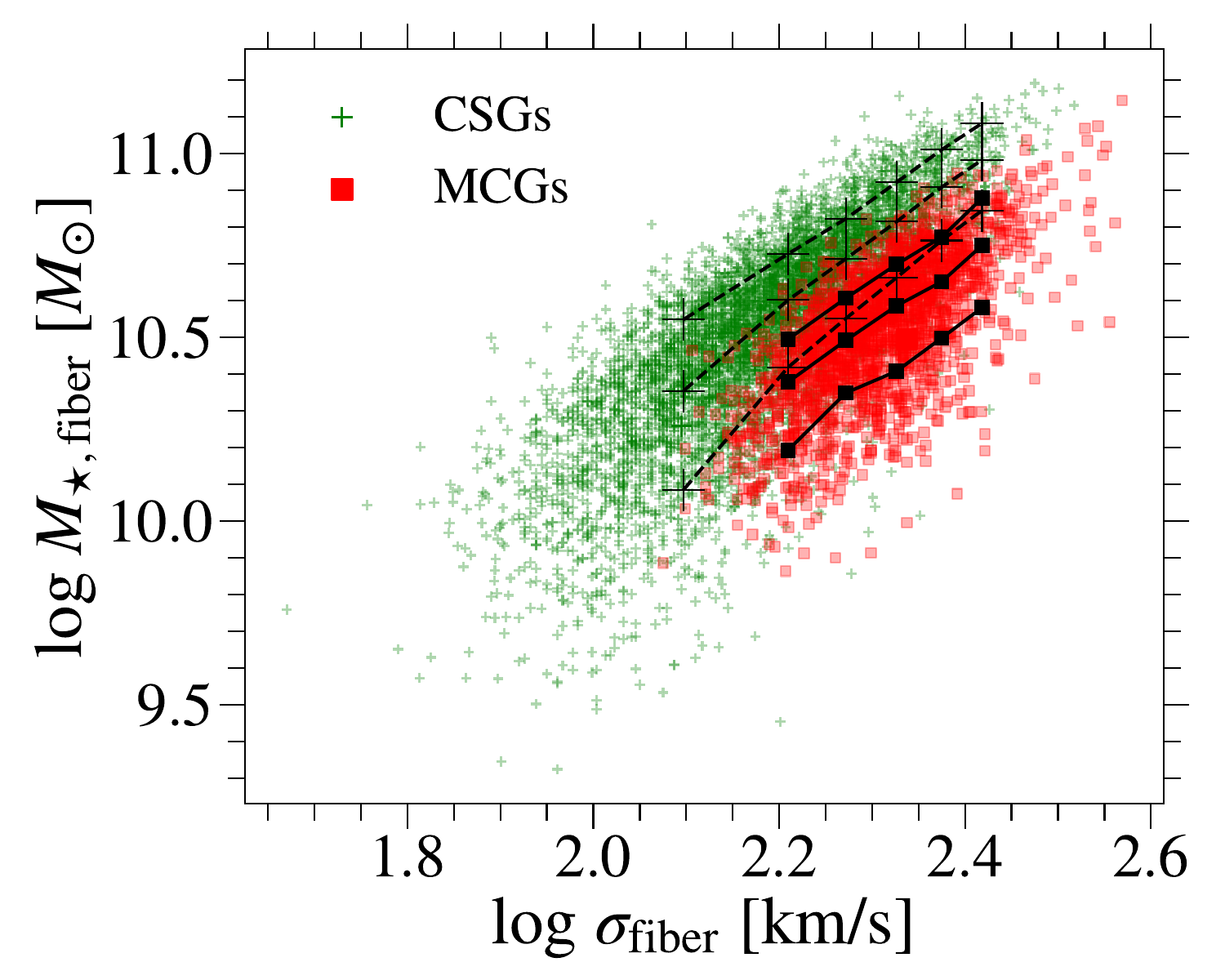}
    \caption{Stellar mass within the SDSS fiber as a function of fiber velocity dispersion for MCGs and CSGs.}
    \label{fig:massa_sigma_fiber}
\end{figure}

In Fig.\, \ref{fig:age_metallicity_alpha_fe_ppxf} we show the mass-weighted age, metallicity and luminosity-weighted [$\alpha$/Fe] distributions as a function of $\sigma_\mathrm{e}$. The values of the 16th, 50th and 84th percentiles for each $\sigma_\mathrm{e}$ bin are shown as black circles and crosses for MCGs and CSGs respectively. Histograms of the stellar population properties for each $\sigma_\mathrm{e}$ bin are shown in appendix\,\ref{statistical_results}. We find similar trends to those in the spectral indices, i.e. MCGs tend to be older, more metal poor and $\alpha$-enhanced than CSGs. We note, however, that for $\sigma_\mathrm{e} > 225$\,km/s there are no statistically significant differences between the ages of MCGs and CSGs and for $\sigma_\mathrm{e} \gtrsim 225$\,km/s differences in [$\alpha$/Fe] are marginal. 

The largest differences between the stellar population properties of MCGs and CSGs are observed in their metallicities. There is a significant number of metal poor MCGs, specially at high $\sigma_\mathrm{e}$, with few to no CSG counterparts. Moreover, both the 16th and 50th percentiles have lower values in MCGs in all bins. The 84th percentiles of the two samples tend to be similar for $\sigma_\mathrm{e} \gtrsim 200$\,km/s, but this could be due to the metallicity limit of the V15 models. Regarding [$\alpha$/Fe], the median and 16th percentile of MCGs and CSGs are similar, but the MCG distribution have a tail of high [$\alpha$/Fe] values. In terms of ages, the most significant difference between the samples is observed in the 16th percentiles, as the 16th percentile of the CSG distribution is shifted to significantly lower values, although this shift decreases with increasing $\sigma_\mathrm{e}$. Basically, this means that while MCGs are predominantly old, CSGs are a mix of young and old galaxies, with the fraction of young galaxies decreasing with increasing $\sigma_\mathrm{e}$.

We note that the two highest $\sigma_\mathrm{e}$ bins are sparsely populated by CGSs while many MCGs lie in these bins. Thus, to test if the similarities in age at $\sigma_\mathrm{e} \geq 225$\,km/s are robust, we built a control sample matched in velocity dispersion and redshift. We find that stellar ages continue to be similar. However, differences in [$\alpha$/Fe] in the $225-250$\,km/s and $\gtrsim 250$\,km/s bins become statistically significant. We note that the median [$\alpha$/Fe] in these bins is similar, the differences lie in the width of the [$\alpha$/Fe] distributions, which is larger in MCGs. However, the increased width is likely to be caused by larger uncertainties in the [$\alpha$/Fe] measurements in MCGs (as they are fainter than CSGs of similar $\sigma_\mathrm{e}$), so we do not considering these differences in [$\alpha$/Fe] meaningful despite their statistical significance. This comparison is shown in appendix\,\ref{sec:sigma_CSGs}.

To explore how age and metallicity correlate at fixed $\sigma_\mathrm{e}$ for MCGs, we plot in the top panel of Fig.\, \ref{fig:stellar_properties_loess} the stellar age as a function of $\sigma_\mathrm{e}$, colouring the points according to their metallicities, and in the bottom panel the metallicity as a function of $\sigma_\mathrm{e}$, colouring the points according to their ages. We used the {\scshape python} implementation of the Locally Weighted Regression (LOESS) method \citep{Cappellari.etal.2013} with a regularisation factor of $f = 0.05$ and a linear local approximation to smooth the colour maps. In the top panel we see that, at fixed $\sigma_\mathrm{e}$, older MCGs tend to have lower metallicity. In the bottom panel we see that there is no clear trend of metallicity with age at fixed $\sigma_\mathrm{e}$, although the youngest and oldest MCGs tend to occupy different regions of the plot. The youngest MCGs have low velocity dispersion and high metallicity, lying in the top left corner. The oldest MCGs are found mainly in two regions: at $\sigma_\mathrm{e} \gtrsim 250$km/s and $[\mathrm{M/H}] \sim 0.4$ and at $ 200\,\mathrm{km/s} \lesssim \sigma_\mathrm{e} \lesssim 300\,\mathrm{km/s}$ and $[\mathrm{M/H}] \sim 0.1$. Note, however, that the age difference between the youngest and oldest MCGs is small ($\lesssim 1$\,Gyr).

In Fig.\,\ref{fig:stellar_properties_loess_stellar_mass} we plot the stellar population properties of MCGs (let panels) and CSGs (right panels) as a function of $\sigma_\mathrm{e}$, colouring each galaxy according to its LOESS smoothed stellar mass to explore if any stellar population property correlates with stellar mass at fixed velocity dispersion. We find that, for MCGs, metallicity increases with stellar mass, while ages and [$\alpha$/Fe] show no trend. On the other hand, CSGs do not show a clear correlation between stellar population properties and stellar mass at fixed $\sigma_\mathrm{e}$. 

Previous works have raised the possibility that a galaxy having larger velocity dispersion at fixed stellar mass might be related to an excess of low mass stars (i.e. a bottom heavy IMF) in their central regions. An accurate measurement of the IMF in quiescent galaxies requires careful modelling of weak absorption features to detect variations of a few percent in absorption strength, something that is only possible with very high S/N data, which is beyond the scope of this work. Yet, it is possible to assess the need for a bottom heavy IMF by comparing dynamical and stellar masses. In Fig.\,\ref{fig:massa_sigma_fiber} we compare the stellar mass within the region cover by the SDSS fiber ($M_{\star,\mathrm{fiber}}$), obtained through spectral synthesis, of MCGs and CSGs at fixed fiber velocity dispersion ($\sigma_\mathrm{fiber}$), a tracer of the dynamical mass within the SDSS fiber. The samples largely overlap, although MCGs have, on average, lower values of $M_{\star,\mathrm{fiber}}$ than CSGs of similar $\sigma_\mathrm{fiber}$.

\subsection{Stellar Population Properties and Environment}
\label{sec:environment}

\begin{figure}
\centering
	\includegraphics[width=\columnwidth, trim=10 40 0 10]{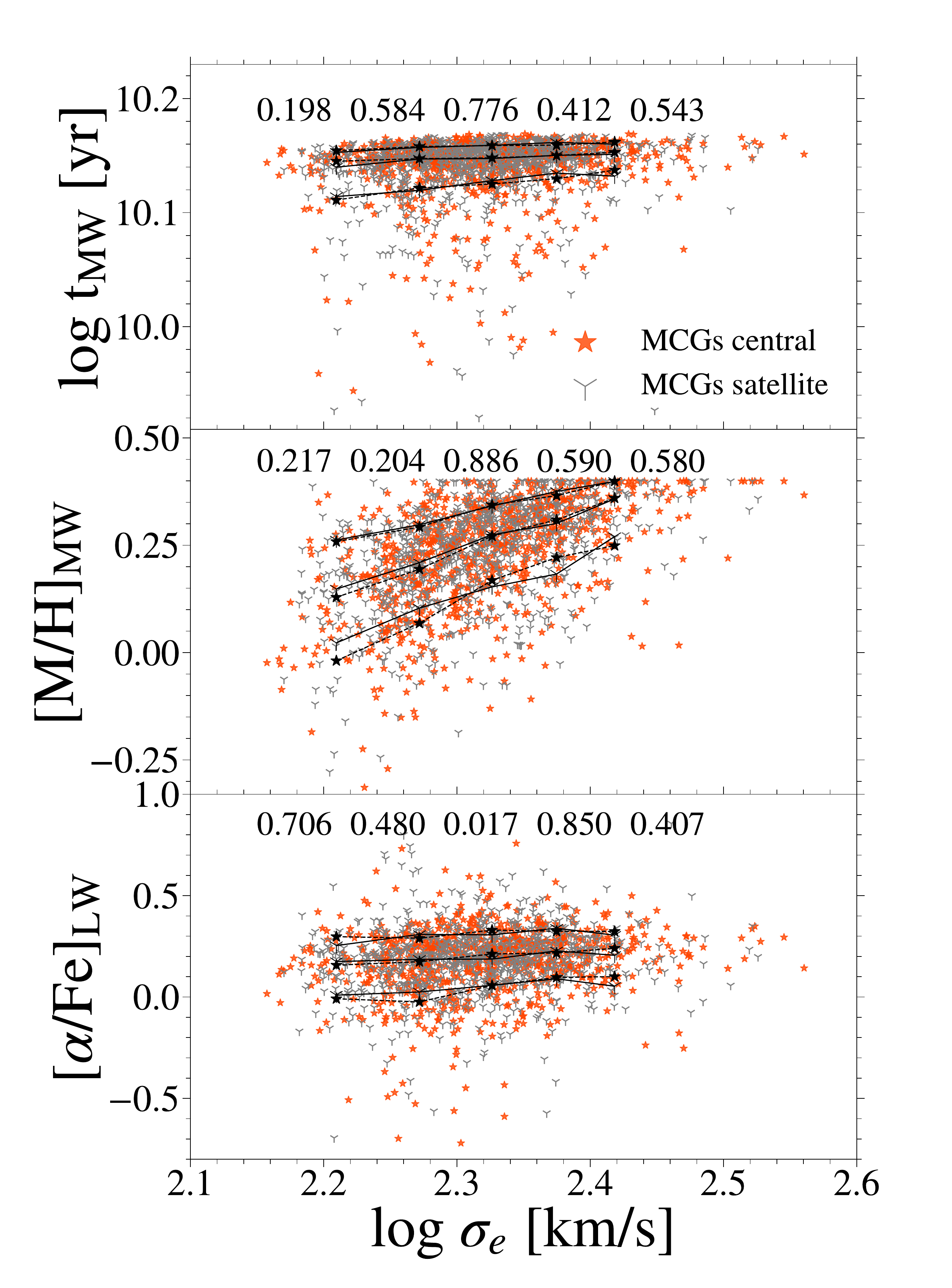}
    \caption{Stellar population properties of central (orange stars) and satellite MCGs (tri-down gray). The star markers connected by dashed lines represent the 16th, 50th and 84th percentiles of the central MCG distributions and the tri-down markers connected by solid lines represent the percentiles of the satellite MCG distributions. The values at the top of each panel are the KS $p-$value tests results for each bin.}
    \label{fig:central_satellite}
\end{figure}

To test if the stellar population properties of MCGs depend on their environment, we divided MCGs in central and satellite galaxies. The central/satellite  classifications were obtained from the \cite{lim17} group catalogue. Of the $1\,858$ MCGs in our sample, $1\,727$  are included in the catalogue. Of these, $856$ are centrals and $871$ satellites.  

In Fig.\, \ref{fig:central_satellite} the stellar population properties as function of $\sigma_\mathrm{e}$ are shown for satellite and central MCGs. A KS test shows that there are no statiscally significant differences between the ages, metallicities and [$\alpha$/Fe] of central and satellites, showing that environmental processes acting exclusively on satellite galaxies do not play a significant role in the formation of MCGs.

\section{Discussion}
\label{sec:discuss}

One complicating factor in interpreting our results is that compact and non-compact galaxies have very different accretion histories. A significant fraction of the stars in present day early-type galaxies were born ex situ, and theoretical studies show that the ex situ fraction depends strongly on stellar mass and galaxy size, with more massive and extended galaxies having larger ex situ fractions \citep{rodriguez-gomez16,clauwens18,tachella19,davison20,pulsoni21}. Thus, we need to understand if the differences between the stellar population properties of MCGs are more likely to reflect differences in their formation pathways or in their post-quenching accretion history. To this end, we need to know what is the expected contribution of stars formed ex situ to the spectra from the region covered by the SDSS fiber. In the following section we will review recent results on the accretion history of early-type galaxies that are pertinent to our work.

\subsection{The accretion history of compact and non-compact early type galaxies}
\label{sec:accretion}

It is not possible to easily distinguish ex situ stars from those born in situ with observations, since their stellar population properties have significant overlap. Because of this, an increasing number of studies have used cosmological simulations to gain insight into the fraction of ex situ stars and their spatial distribution. These theoretical studies have found that the ex situ fraction inside a given radius increases with galactocentric distance, as stars formed in situ tend to be concentrated close to their original formation sites \citep{cook16,rodriguez-gomez16,pulsoni21}. Simulations show that the accretion history of compact and average-sized galaxies is different. Thus, we will discuss these two cases separately.

In the case of average-sized galaxies, ex situ fractions within 1\,$R_\mathrm{e}$ are small for galaxies with $\log M_\star/M_{\odot} \lesssim 10.5$, while they steeply increase for larger masses, reaching $\sim 40\%$ in galaxies with $10.5 \lesssim \log M_\star/M_{\odot} \lesssim 11.0$, and being above $> 50\%$ for galaxies with $\log M_\star/M_{\odot} \gtrsim 11.0$ \citep{tachella19}. In the most massive galaxies, stars accreted in major mergers dominate, while for galaxies in the mass range $10.5 \lesssim \log M_\star/M_{\odot} \lesssim 11.0$ stars accreted in major and minor mergers contribute equally \citep{tachella19}. Ex situ stars are deposited at larger galactocentric distances in order of decreasing merger mass ratio \citep{rodriguez-gomez16}, meaning that stars accreted in major mergers dominate the ex situ fraction in inner regions.

For CSGs with $\sigma_\mathrm{e} \gtrsim 150$\,km/s (corresponding to $\log M_\star/M_{\odot} \gtrsim 10.7$), the SDSS fiber typically probes a radius of $\sim 0.2-1.0\,R_\mathrm{e}$ (see Fig. \, \ref{fig:fibra}). In this case, given the results described above, we expect a significant contribution of ex situ stars accreted in major merges to the observed spectra. However, as it is not expected that two merging galaxies of similar mass have very different stellar populations, a larger contribution of stars accreted in major mergers cannot be the driving factor of the differences in stellar populations properties of MCGs and CSGs.

\citet{davison20} studied the ex situ fraction of compact galaxies (defined as those with sizes within the 33rd percentile in
a given bin of stellar mass) in the EAGLE simulation, finding that  within  1\,$R_\mathrm{e}$ they are below 10\% up to $\log M_\star/M_{\odot} \sim 10.7$. Even within $2\,R_\mathrm{e} < R < 100\,$kpc the ex situ fractions are low, being $\sim 20\%$ for galaxies in this mass range. For larger masses, the ex situ fractions quickly increase, reaching $\sim 30\%$ within 1\,$R_\mathrm{e}$ and $\sim 40\%$ within 2\,$R_\mathrm{e}$ at $\log M_\star/M_{\odot} \sim 11.0$. These ex situ fractions include both stars accreted before and after a galaxy becomes quiescent, however. In a recent work, \citet{lohmann23} used the Illustris TNG100 simulation to study the merger history of a sample of $z = 0$ compact galaxies, selected using similar criteria as those in our work. They found that $\sim 70\%$ of the galaxies in their sample have not experienced either major or minor mergers after quenching their star formation activity. Observations of compact quiescent galaxies are in agreement with the theoretical results: \citet{Schnorr.et.al.2021} argued that MCGs in the Mapping Nearby Galaxies at APO (MaNGA) survey experienced a quiet accretion history after quenching, based on their stellar kinematics and stellar population properties. Given the discussion above, we do not expect a significant contribution from stars accreted in dry mergers to the spectra of the majority of MCGs.

In summary, we conclude that while MCGs and CSGs are expected to have different accretion histories, this is unlikely to be the main driver of the differences in their stellar population properties.

\subsection{The Origins of the Differences in Stellar Population Properties between MCGs and CSGs}
\label{pops_MCG_CSG}

Our results show that the stellar population properties of MCGs differ significantly from those of CSGs with similar $\sigma_\mathrm{e}$. The age and [$\alpha$/Fe] of MCGs show no significant variation with $\sigma_\mathrm{e}$; MCGs are predominantly old galaxies with a narrow age distribution, [$\alpha$/Fe] varying from solar to super-solar. In contrast, CSGs have a broad age distribution, with the median age increasing with $\sigma_\mathrm{e}$, while the width of the distribution decreases. Their [$\alpha$/Fe] ratio mildly increasing with $\sigma_\mathrm{e}$. Lastly, MCGs consistently have lower metallicity than CSGs.

\subsubsection{Age and [$\alpha$/Fe]}
\label{sec:stellar_age}

When comparing the differences in age and [$\alpha$/Fe], one notices that both follow a similar pattern.  For $\sigma_\mathrm{e} \lesssim 225$\,km/s (from here on low-$\sigma_\mathrm{e}$), CSGs have lower median ages and [$\alpha$/Fe] than MCGs (but note that there still is some overlap between the samples). For $\sigma_\mathrm{e} \gtrsim 225$\,km/s (high-$\sigma_\mathrm{e}$) there are no longer any significant differences. We argue that the age and [$\alpha$/Fe] differences at low-$\sigma_\mathrm{e}$ are due to a progenitor bias effect, and we lay out our reasoning below.

Previous works studying the size-age relation of quiescent galaxies \citep{fagioli16,damjanov19} found that, below $\log M_\star/M_{\odot} \sim 11$, both the size growth of individual galaxies and progenitor bias effects are important for the evolution of the mean size of quiescent galaxies, thus average sized quiescent galaxies in this mass range should present a variety of ages. On the other hand, at larger masses the merger driven growth of individual galaxies plays a significant role, meaning that older galaxies dominate at high masses. Considering the mean $\sigma_\mathrm{e}$ of CSGs with $\log M_\star/M_{\odot} > 11$ is $\sigma_\mathrm{e} \sim 200$\,km/s, these results are consistent with the old ages and supersolar $[\alpha$/Fe] of high-$\sigma_\mathrm{e}$ CSGs and with the broad age and [$\alpha$/Fe] distribution of low-$\sigma_\mathrm{e}$ CSGs.

\subsubsection{Stellar Mass within the SDSS fiber}
\label{sec:mass_fiber}

MCGs have, on average, lower stellar masses within the region covered by the SDSS fiber than CSGs of similar $\sigma_\mathrm{fiber}$ (see Fig.\,\ref{fig:massa_sigma_fiber}). Considering that $\sigma_\mathrm{fiber}$ is a tracer of the dynamical mass, this suggests that the difference between the dynamical and stellar mass in the inner few kpc tends to be larger in MCGs, implying a bottom heavier IMF and/or a higher dark matter fraction (in appendix \ref{sec:sigma_1kpc} we show that a similar conclusion is reached when comparing the central stellar surface density within $1$ kpc at fixed $\sigma_\mathrm{1kpc}$). Since we cannot rule out nor assess the likelihood of either possibility based only on the results presented in this paper, we defer a detailed investigation of this question to future work, limiting ourselves to briefly discuss their plausibility below.

Studies of the stellar population properties of nearby compact quiescent galaxies point to them having an enhanced fraction of low mass stars up to $\sim 2R_\mathrm{e}$ \citep{martin-navarro15,ferre-mateu17}. Moreover, recent work have suggested that old compact galaxies have a bottom-heavier IMF slope than quiescent galaxies of similar $\sigma_\mathrm{e}$ \citep{martin-navarro23}. We note that an increased fraction of low mass stars would result in fewer supernovae overall, decreasing [$\alpha$/Fe] as well as the stellar metallicity \citep{martin-navarro16}. This is inconsistent with MCGs having similar [$\alpha$/Fe] to high-$\sigma_\mathrm{e}$ CSGs. However, a time-dependent IMF, switching from top to bottom-heavy, can simultaneously account for both high [$\alpha$/Fe] values and an excess of low masses stars, thus being consistent with our results \citep{deMasi19, denBrok.etal.2024}.

Simulations of compact galaxy formation predict low dark matter fractions within $R_\mathrm{e}$ in these objects \citep{lapiner23}. Dynamical modelling of the stellar kinematics also favour low dark matter fractions, with \citet{yildirim17} estimating a median dark matter fraction within $R_\mathrm{e}$ of just $11\%$ for a sample of 15 nearby compact quiescent galaxies. In contrast to these studies, \citet{buote19} estimated a dark matter fraction within $R_\mathrm{e}$ of $f_{\mathrm{DM}} \sim 40\%$ for Mrk\,1216, a nearby compact galaxy ($\log M_\star/M_{\odot} \sim 11.3$, $R_\mathrm{e} = 3.0$\,kpc, $\sigma_\mathrm{e} = 308$\,km/s; \citealt{yildirim17}), based on a hydrostatic equilibrium analysis of the X-ray plasma emission, so there are exceptions although how common they are is yet to be determined.

\subsubsection{Stellar Metallicity}
\label{sec:metal}

What is the origin of the differences in metallicity between the MCG and CSG samples at fixed $\sigma_\mathrm{e}$? They seem to have a different origin from the age and [$\alpha$/Fe] differences, as MCGs have lower metallicities in all $\sigma_\mathrm{e}$ bins. Since at fixed $\sigma_\mathrm{e}$ the metallicity and stellar mass of MCGs are positively correlated (which is not the case in CSGs, see Fig.\,\ref{fig:stellar_properties_loess_stellar_mass}), one might wonder if the differences in metallicity are related to MCGs having lower $M_{\star,\mathrm{fiber}}$. It is possible, since both a bottom heavier IMF and a higher dark matter fraction within the region sampled by the SDSS spectra are consistent with lower metallicities. A bottom heavier IMF would lead to fewer type II SNe, while a higher dark matter fraction implies lower stellar masses and, as the stellar mass is proportional to the integral of metal production \citep{baker_maiolino23}, a lower metallicity. Alternatively, it has been proposed that the global stellar metallicity is better correlated with the ratio $M_\star/R_\mathrm{e}$ \citep{barone18}, which is consistent with MCGs having lower metallicity at fixed $\sigma_\mathrm{e}$ since this ratio is larger in CSGs, as the difference in mass at fixed velocity dispersion between the samples is larger than the difference in size (CSGs are $\sim 0.7$\,dex more massive than MCGs of similar $\sigma_\mathrm{e}$ but only about $0.4$\,dex larger).

A third possibility is that the differences in metallicity are due to metallicity gradients. As shown in Fig.\,\ref{fig:fibra}, the SDSS fiber typically covers $\sim 1-2R_\mathrm{e}$ in MCGs, while covering $\lesssim 1R_\mathrm{e}$ in CSGs. Consequently, in the presence of a negative metallicity gradient, metallicities in MCGs would be biased to lower values. Since mass and size are correlated, one would expect that, at fixed velocity dispersion, lower metallicities would be measured in lower mass galaxies. To assess the magnitude of this effect, we use integral field spectroscopic observations of 39\,MCGs which were observed as part of the MaNGA survey, measuring the stellar metallicity in circular apertures of increasingly larger radii, from $0.25$ to $1.5R_\mathrm{e}$. We find that, on average, the metallicity measured in an aperture of $1.5R_\mathrm{e}$ is 0.06\,dex lower than the one measured in a $0.25R_\mathrm{e}$ aperture, but there is significant variation between different galaxies (see appendix \ref{sec:mcgs_gradients} for details). While the trends in this small sample might not necessarily reflect those in the full sample, this test shows that one should be cautious in interpreting the differences between the metallicity of MCGs and CSGs as it is possible that metallicity gradients account for most, if not all of them. In conclusion, distinguishing between the three possibilities presented here requires spatially resolved spectroscopic observations.

\subsection{Comparison with Previous Works}
\label{MCG_Manga}

\begin{figure}
\centering
	\includegraphics[width=\columnwidth, trim=10 40 0 10]{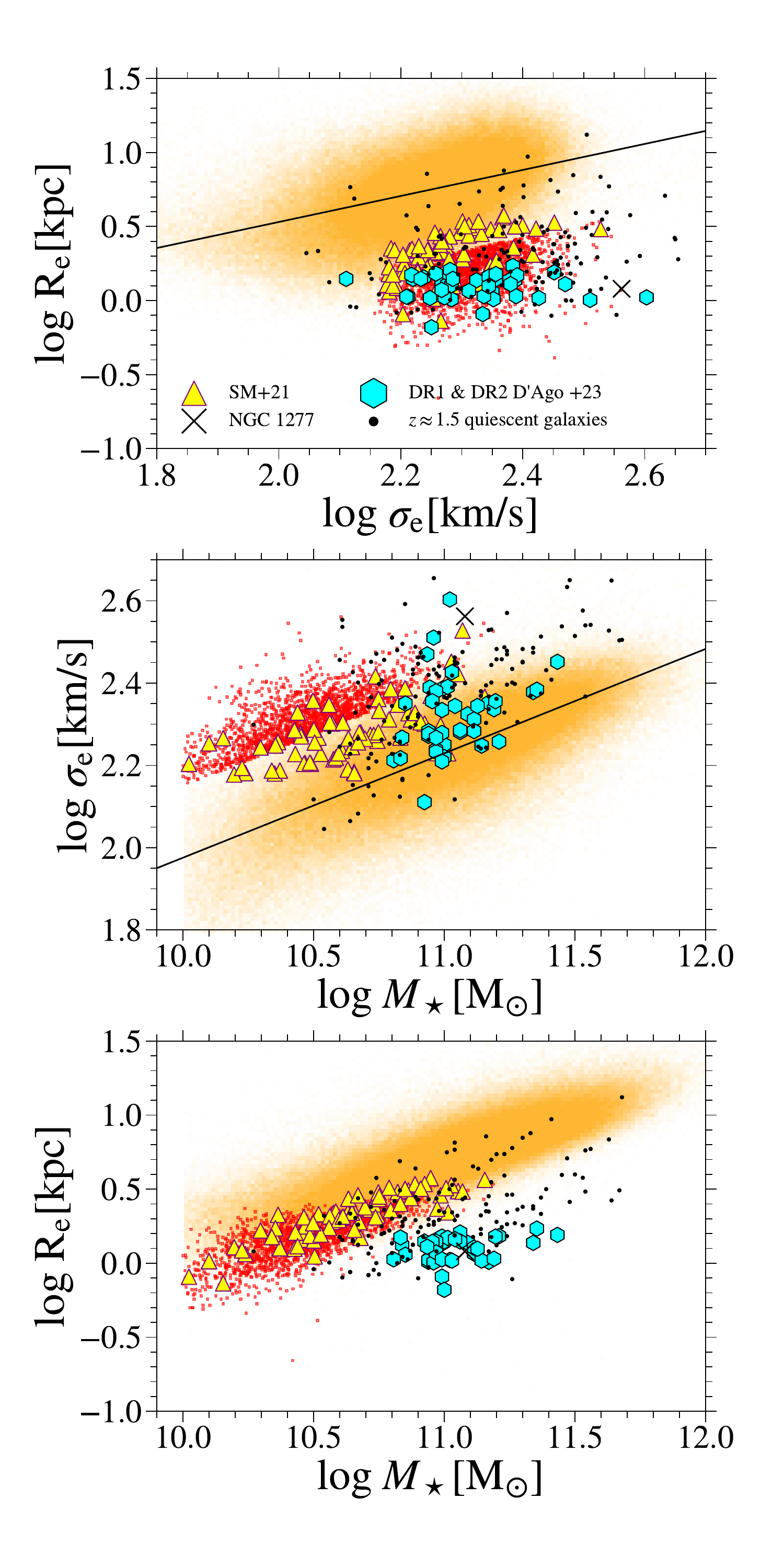}
    \caption{Similar to Fig.\, \ref{fig:final_sample}. Top panel: log $R_e$ vs. log $\sigma_\mathrm{e}$. Middle panel: log $\sigma_\mathrm{e}$ vs. log $M_{ \star}$.
    Bottom panel: log $R_e$ vs. log $M_{ \star}$.  SDSS quiescent galaxies are shown in orange, MaNGA galaxies are shown as yellow triangles, INSPIRE galaxies as cyan octagons, NGC\,1277 as an "$\mathrm{x}$", and $z \approx 1.5$ quiescent galaxies as black points.}
    \label{fig:MCG_sdss_manga_relic}
\end{figure} 

 \begin{figure}
 \centering
	\includegraphics[width=\columnwidth, trim=10 40 0 10]{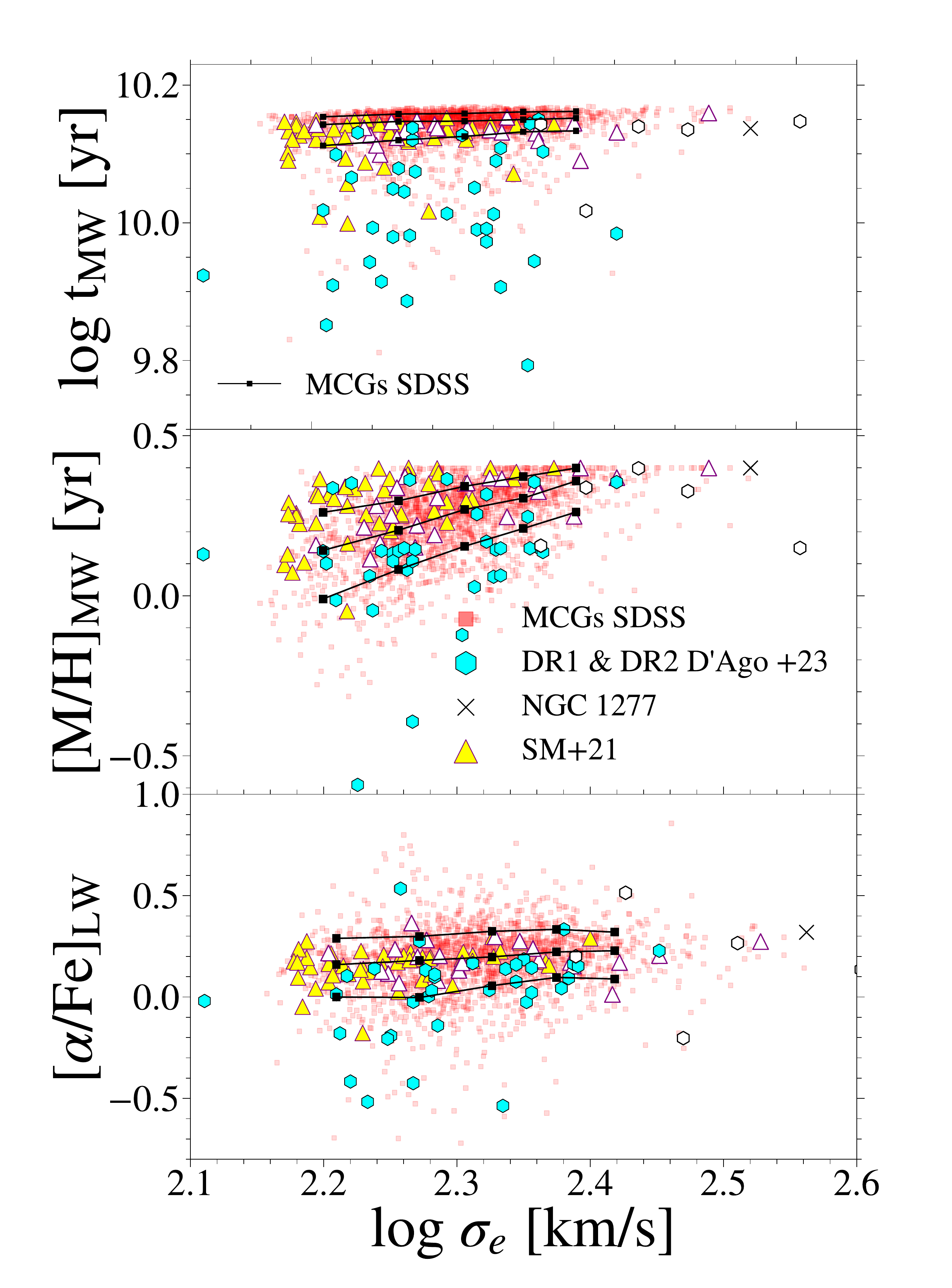}
    \caption{Stellar population properties as a function of $\sigma_\mathrm{e}$ for SDSS MCGs (red squares), MaNGA galaxies ( yellow and white triangles), INSPIRE galaxies (cyan and white hexagons) and the relic galaxy NGC $1277$ (marked with an "$\mathrm{x}$"). The white triangles and hexagons correspond to MaNGA and INSPIRE galaxies that are in the same region of the log$\sigma_e$-log$M_{star}$ diagram as MCGs. The square markers connected by solid lines are the 16th, 50th and 84th percentiles of the SDSS MCG distributions.}
    \label{fig:MCG_sdss_manga_stellar_pro}
\end{figure}

Studies of low redshift compact quiescent galaxies have been mostly limited to small samples of high mass ($\log M_\star/M_{\odot} \gtrsim 10.8$) extremely compact ($R_\mathrm{e} \lesssim 2$\,kpc) galaxies \citep{Trujillo.etal.2014,ferre-mateu17,yildirim17,Buitrago.et.al.2018,Spiniello.etal.2021b}. A subset of these galaxies have been shown to have formed most of their mass by $z \sim 2$ and to have had a quiet accretion history since then, these are called relic galaxies \citep{beasley18}. In this section we compare the stellar population properties of MCGs to those of the nearby relic galaxy NGC\,$1277$, the $z \sim 0.3$ compact quiescent galaxy sample from \citet{Spiniello.et.al.2023} and the local compact quiescent galaxy sample of \cite{Schnorr.et.al.2021}. The stellar population properties of NGC\,1277 and the \cite{Schnorr.et.al.2021} compact galaxy sample (hereafter the SM+21) were measured from their SDSS spectra following the methodology outlined in Sec.\,\ref{sec:spectral_synthesis}. The stellar population properties of the \citet{Spiniello.et.al.2023} sample were measured from publicly available X-SHOOTER spectra, released as part of the INvestigating Stellar Population In RElics \citep[INSPIRE]{Spiniello.et.al.2021} second data release (hereafter the INSPIRE sample).

In Fig.\,\ref{fig:MCG_sdss_manga_relic} we show the distribution of MCGs, the MaNGA sample and the INSPIRE sample in the log $R_\mathrm{e}$ vs. log $\sigma_\mathrm{e}$, log $M_{\star}$ vs. log $\sigma_\mathrm{e}$ and log $R_\mathrm{e}$ vs. log $M_{\star}$ diagrams. We note that these samples were selected using different criteria. Compact galaxies in the MaNGA sample were selected as those with sizes $1\sigma$ below the median in a given $\sigma_\mathrm{e}$ bin (25\,km/s bin width, with $\sigma_\mathrm{e} = 150-350$\,km/s), while compact galaxies in the INSPIRE sample satisfy the criteria $\log M_\star/M_{\odot} \geq 10.9$ and $R_\mathrm{e} \leq 1.5$\,kpc. For comparison, we show as black dots a sample of quiescent galaxies in the redshift range $1 \lesssim z \lesssim 2.5$, built by combining galaxies extracted from \citet{VandeSande.etal.2013}, \citet{Belli.etal.2014}, \citet{Gargiulo.etal.2015} and \citet{Belli.etal.2017}. Note that, besides being shifted to smaller sizes at fixed stellar mass compared to SDSS quiescent galaxies, galaxies in the high redshift sample are also shifted to higher velocity dispersion. In the log $R_\mathrm{e}$ vs. log $M_{\star}$ diagram, MCGs and the MaNGA sample occupy similar regions, but MaNGA galaxies have lower $\sigma_\mathrm{e}$ at fixed stellar mass. Meanwhile, galaxies in the INSPIRE sample are the most compact, but they are spread out in the log $M_{\star}$ vs. log $\sigma_\mathrm{e}$ diagram.  

In Fig.\,\ref{fig:MCG_sdss_manga_stellar_pro} we compare the stellar population properties of MCGs with those of NGC1277 and the MaNGA and INSPIRE samples. MaNGA and INSPIRE galaxies that are $+2\sigma$ outliers in the  $\log M_{\star}$ vs.  $\log \sigma_\mathrm{e}$ relation are plotted as empty symbols. INSPIRE galaxies, on average, have younger ages and lower [$\alpha$/Fe] than MaNGA galaxies and MCGs, despite being significantly more compact. This is not surprising, since previous studies have reported that samples of compact quiescent  galaxies selected using criteria based solely on stellar mass and size are comprised of objects with a variety of ages \citep{ferre-mateu12,Buitrago.et.al.2018}. In regards to MaNGA galaxies, most have mass-weighted ages of $\sim 12$\,Gyr, similar to MCGs, although the fraction of younger galaxies is larger in the MaNGA sample. The [$\alpha$/Fe] distribution of both samples is also similar. On the other hand, MaNGA galaxies tend to have larger metallicities than MCGs. Lastly, we find that outliers in the  $\log M_{\star}$ vs. $\log \sigma_\mathrm{e}$ relation in both the MaNGA and INSPIRE samples tend to be old, while non-outliers have a wide range of ages. Outliers in the MaNGA sample also have, on average, lower metallicities than non-outliers. In contrast, no clear trend is seen in the INSPIRE sample. 

In conclusion, the results discussed above highlight the fact that compact quiescent galaxies are not a homogeneous population, showing significant spread in their stellar population properties at fixed velocity dispersion. Additionally, they imply that the position of a compact galaxy in the $\log M_{\star}$ vs. $\log \sigma_\mathrm{e}$ diagram is a good age predictor, in the sense that galaxies with higher velocity dispersion at fixed stellar mass tend to be older.

\subsection{Insights on the Formation of MCGs}
\label{sec:formation}

We start this section by noting that since it is very difficult to measure ages with precision for systems older than $\gtrsim 9$\,Gyr, as evolution in the isochrones is very slow at late times \citep{conroy13}, therefore the very old mass-weighted ages ($\sim13$\,Gyr) we measure for MCGs should not be taken at face value. Nonetheless, while we cannot determine their ages with high precision, a lookback time of $\sim 9$\,Gyr corresponds to $z \sim 1.4$, so it is reasonable to conclude that MCGs quenched at high redshifts. This raises the question: are they passively evolving? As previously discussed in Sec.\,\ref{sec:accretion}, both simulations and observations point to $z \sim 0$ compact quiescent galaxies in general having a quiet accretion history, so it is reasonable to assume that most MCGs are the passively evolving descendants of high redshift quiescent galaxies. We will test this assumption in future work, but for the following discussion we assume it is true.

\subsubsection{The Progenitors of MCGs and High-$\sigma_\mathrm{e}$ CSGs} 
\label{progenitors}

According to the two-phase formation scenario, compact quiescent galaxies grow in size by the accretion of several lower-mass galaxies which form an extended envelope around the compact core \citep{Naab.etal.2009,Oser.etal.2010,hilz12}. In this scenario, stars belonging to the high-$z$ quiescent progenitor are expected to dominate in the inner $1-2$\,kpc \citep{hopkins09b,bezanson09}. Therefore, considering that we interpreted high-$\sigma_\mathrm{e}$ CSGs as being the descendants of old quiescent galaxies that grew by dry mergers, and taking into account that the SDSS fiber covers between $\sim 0.6-2.8$\,kpc, when comparing the stellar population properties of MCGs and high-$\sigma_\mathrm{e}$ CSGs we learn about possible differences between their high redshift progenitors. In this regard, while the similarities in both age and [$\alpha$/Fe] point to the progenitors of both samples quenching at the same epoch, MCGs having, on average, lower stellar masses within the region covered by the SDSS fiber than CSGs of similar $\sigma_\mathrm{fiber}$ coupled with the differences in stellar metallicity suggests that, despite largely overlapping, the progenitor populations of MCGs and high-$\sigma_\mathrm{e}$ CSGs are not the same. In particular, it is notable that galaxies with very high $M_\mathrm{\star,fiber}$ (i.e. very high stellar mass surface density) are absent among MCGs (see Fig.\,\ref{fig:massa_sigma_fiber}). A similar conclusion can be reached by investigating Fig.\,\ref{fig:MCG_HST} and Fig.\,\ref{fig:MCG_sdss_manga_relic}. While there are plenty of quiescent galaxies with similar sizes to MCGs at $z \gtrsim 1.5$, MCGs are larger than typical high-$z$ quiescent galaxies, so even if they are passively evolving, MCGs cannot be considered representative of the high-$z$ quiescent population.

Recent work have reported that the evolution of the half-mass radius with redshift is much weaker than the evolution of the half light radius, raising the possibility of a much smaller contribution of dry mergers to the size growth of quiescent galaxies than usually assumed \citep{suess19,ibarra22,avila23}. In this case, our conclusion that MCGs and high-$\sigma_\mathrm{e}$ CSGs descend from different populations of galaxies that quenched roughly at the same epoch is still valid, as the observed differences in size and stellar mass at fixed effective velocity dispersion would be largely already in place before the onset of quenching in this scenario.

\subsubsection{The Formation of High Redshift Quiescent Galaxies} 
\label{formation}

Studies on the formation of high redshift quiescent galaxies can give us insight on how MCGs formed. Recent studies of $z \gtrsim 1$ compact quiescent galaxies have proposed that these objects can follow two paths to quiescence: one associated to a fast quenching of star formation and structural change, and other where no structural changes occur and there is a diversity of quenching timescales \citep{belli19,suess21,tacchella22}. Fast quenching and structural change are believed to be related to galaxy mergers and interactions, as galaxies with short quenching timescales are predominantly central galaxies in overdense environments \citep{belli19}. 

High-resolution hydrodynamical simulations offer a clue on how structural change followed by star formation quenching might unfold: wet mergers and interactions can efficiently drive large amounts of gas to the centre of galaxies, igniting a central starburst and turning a more extended galaxy compact \citep{dekel14,zolotov15,lapiner23}. After compaction, a combination of gas exhaustion and supernova feedback quenches the galaxy, with recurrent AGN feedback heating the circumgalactic gas and keeping the galaxy quenched  \citep{lapiner21}. Observational studies have reported that at $z \gtrsim 1$ there exists a sizeable population of star-forming galaxies with compact star forming components embedded in a larger stellar structure \citep{puglisi19,magnelli23}, which could represent a post-compaction population. As these galaxies have low gas fractions \citep{franco20,gomes-guijarro19,gomes-guijarro22} they will soon quench, thus offering tentative evidence of a link between compaction events and the formation of compact quiescent galaxies.

On the other hand, in cosmological simulations, many compact quiescent galaxies do not experience a compaction event. These descend from compact star-forming galaxies that assembled most of their mass early, when the Universe was denser, and grew little in size while their stellar mass increased \citep{wellons15,lohmann23}. Simulated galaxies following this path do not go through a structural transformation. It has recently been reported that some massive star-forming galaxies at $z \sim 8$ have stellar densities in the inner $\sim 100$\,pc comparable to those of $z \sim 2$ compact quiescent galaxies \citep{baker23,baggen23}, providing further evidence that some galaxies did build their cores very early in the history of the Universe.

Coming back to MCGs, we note that, while we cannot estimate their quenching timescales with a degree of precision that allows us to assess the prevalence of these formation pathways, the environment of MCGs can give us some clues. As we have discussed in Sec.\,\ref{sec:environment}, a large fraction of MCGs are satellites, moreover, those that are centrals tend to inhabit low density environments (see \citealt{Schnorr.et.al.2021}). As previously mentioned, $z \sim 2$ quiescent galaxies with short quenching timescales are predominantly central in overdense environments, so we speculate that it is more likely that MCGs have followed the early formation pathway.

\subsubsection{Enhanced Black Hole Growth and Galaxy Quenching} 

Both simulations and observations point to gas depletion as a cause of quenching in the high redshift Universe \citep{lapiner21,whitaker21}. What halts the inflow of gas to galaxies, however, is still an open question. It has been proposed that shock heating of gas to the virial temperature upon infall into a massive dark matter halo can effectively halt gas supply to the galaxy triggering quenching \citep{dekel06}. Alternatively, it has been suggested that heating of the circumgalactic medium by AGN feedback is the main quenching mechanism acting on central galaxies \citep{piotrowska22}. This is supported by studies of central galaxies at low redshifts, which have found that the central stellar velocity dispersion, that tightly correlates with SMBH mass, is the most predictive parameter of quenching \citep{brownson22,bluck22}.

Observations of high-$z$ star-forming galaxies point to a connection between elevated SMBH growth and galaxy compactness \citep{kocevski17,aird22}, interpreted as a consequence of a relation between SMBH growth and the gas density within the central 1\,kpc \citep{ni21,li21,yang22}. So far, only a few compact quiescent galaxies have had their SMBH mass measured dynamically, but the available evidence points to their SMBH masses being outliers in the $M_\mathrm{BH}-L_\mathrm{Bulge}$ relation, while being consistent with the local $M_\mathrm{BH} - \sigma_\mathrm{bulge}$ relation \citep{yildirim15,walsh15,walsh16,walsh17,cohn23}, implying that compact quiescent galaxies have larger SMBH masses than galaxies of similar stellar mass. If the integrated history of the AGN feedback (which is encoded in the SMBH mass) is what determines when a galaxy quenches \citep{piotrowska22,bluck23}, then galaxies with larger stellar velocity dispersion at fixed stellar mass were quenched earlier, which is consistent with the uniformly old ages of MCGs.

\section{SUMMARY AND CONCLUSIONS}
\label{sec:conc}

We investigated the stellar population properties of a sample of massive compact quiescent galaxies and of a control sample of average-sized quiescent galaxies matched in redshift using spectral indices and stellar population synthesis. Our main results are:

\begin{itemize}
    \item The analyses of spectral indices and stellar population synthesis lead to similar conclusions: at low-$\sigma_\mathrm{e}$ ($\lesssim 225$\,km/s), MCGs are older, more metal-poor and $\alpha$-enhanced than CSGs. At high-$\sigma_\mathrm{e}$ ($\gtrsim 225$\,km/s),  MCGs and CSGs have similar mass-weighted age and $\alpha$-enhancement, but MCGs have lower metallicities;  
    \item We interpret the differences in age at low-$\sigma_\mathrm{e}$ as due to a progenitor bias effect. Low-$\sigma_\mathrm{e}$ CSGs have a variety of ages. In constrast, MCGs are predominantly old;
    \item  The origin of the differences in metallicity is not clear. We suggest three possibilities: MCGs having lower stellar masses within the region probed by the SDSS spectra, MCGs having a bottom heavier IMF or metallicity gradients;
    \item We found that MCGs have lower $\log M_{\star, \mathrm{fiber}}$ than CSGs at fixed $\sigma_{\mathrm{fiber}}$. Considering that the $\sigma_{\mathrm{fiber}}$ is an observable related to dynamical mass, this suggests MCGs tend to have a higher dark matter fractions within the inner few kiloparsecs and/or a bottom heavier IMF;
    \item We found that the position in the $\log M_{\star}$ vs. $\log \sigma_{\mathrm{e}}$ plane is a good age predictor: compact galaxies with higher $\sigma_e$ at fixed $M_{\star}$ tend to be older;
    \item We compared the stellar population properties of central and satellite MCGs and we found no statistically significant differences. This suggests that environmental processes exclusively acting on satellite galaxies do not play a significant role in the formation of MCGs.

\end{itemize}

\section*{Acknowledgements}
The authors thank the anonymous referee for their comments and suggestions, which led to an improved version of the manuscript. KSC acknowledges the coordination for the improvement of higher education personnel (CAPES) for the financial support (88887.629089/2021-00). ASM acknowledges the financial support from the Brazilian national council for scientific and technological development (CNPq). TVR also acknowledges CNPq for funding through grant number 304584/2022-3. MT thanks the support of CNPq (process 312541/2021-0).
\section*{Data Availability}

The SDSS data release 14 data are available at  \url{https://data.sdss.org/sas/dr14}. The data generated by this work will be shared upon request to the corresponding author. 



\bibliographystyle{mnras}
\bibliography{example} 




\appendix

\section{Inclination effects}
\label{ellipticity_effects}

\begin{figure}
	\includegraphics[width=\columnwidth, trim=10 40 0 10]{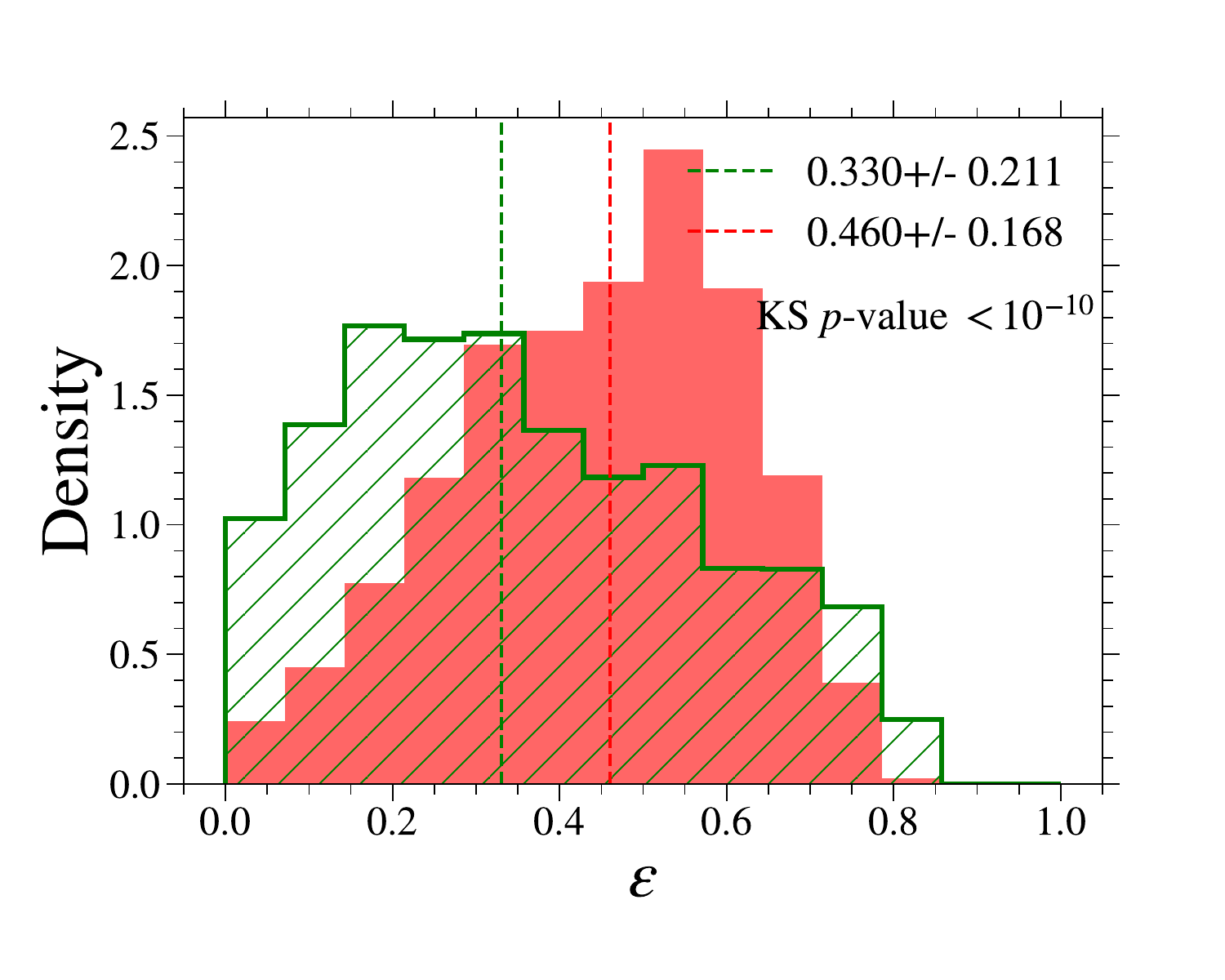}
    \caption{Distribution of ellipticity of MCGs and CSGs.}
    \label{fig:ellipticity_MCGs_CSGs}
\end{figure}

\begin{figure*}
\centering
	\includegraphics[width=\textwidth]{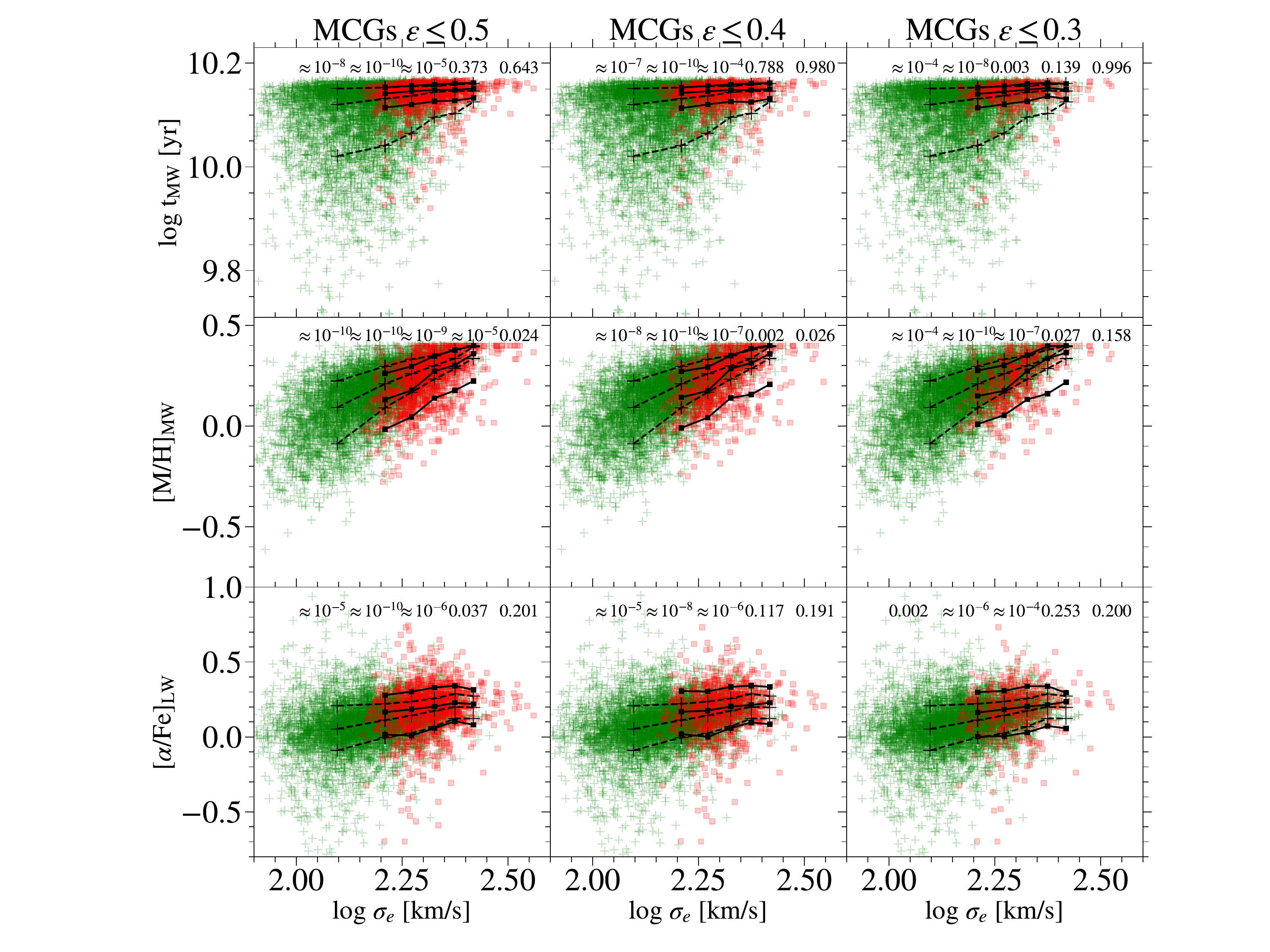}
    \caption{Similar to Fig.\, \ref{fig:age_metallicity_alpha_fe_ppxf}. In each column we selected a subgroup of MCGs according to their ellipticity. Left column: MCGs with $\epsilon \leq 0.5$, middle column: MCGs $\epsilon \leq 0.4$ and right column: $\epsilon \leq 0.3$.}
   \label{fig:stellar_population_properties_ellipticity}
\end{figure*}

Dynamical analyses of fast-rotating early type galaxies have shown that typically $\sigma_\mathrm{z} < \sigma_{R}$ \citep{cappellari16}, meaning that our MCG selection criteria is biased against face-on galaxies, as can be seen in Fig. \, \ref{fig:ellipticity_MCGs_CSGs}. To test if this inclination bias can affect our results, in Fig. \, \ref{fig:stellar_population_properties_ellipticity} we compared the stellar population properties of CSGs to sub-samples of MCGs, including only galaxies with ellipticity below a certain threshold. In the case of $\epsilon \leq 0.5$ and $\epsilon \leq 0.4$, the results are consistent with the full sample shown in Fig. \, \ref{fig:age_metallicity_alpha_fe_ppxf}. For $\epsilon \leq 0.3$, differences in metallicity in the highest $\sigma_\mathrm{e}$ bin are no longer significant. However, we note that this criterion is very restrictive and remove a significant number of MCGs, such that for $\sigma_e \geq 250$ km/s only approximately $100$ MCGs remain in this subsample. Therefore, it is possible that the differences are no longer significant simply because the number of galaxies in this bin is very small.

\section{MCGs and the Stellar mass-size relation}
\label{sec:mass-size}

\begin{figure}
	\includegraphics[width=\columnwidth, trim=10 40 0 10]{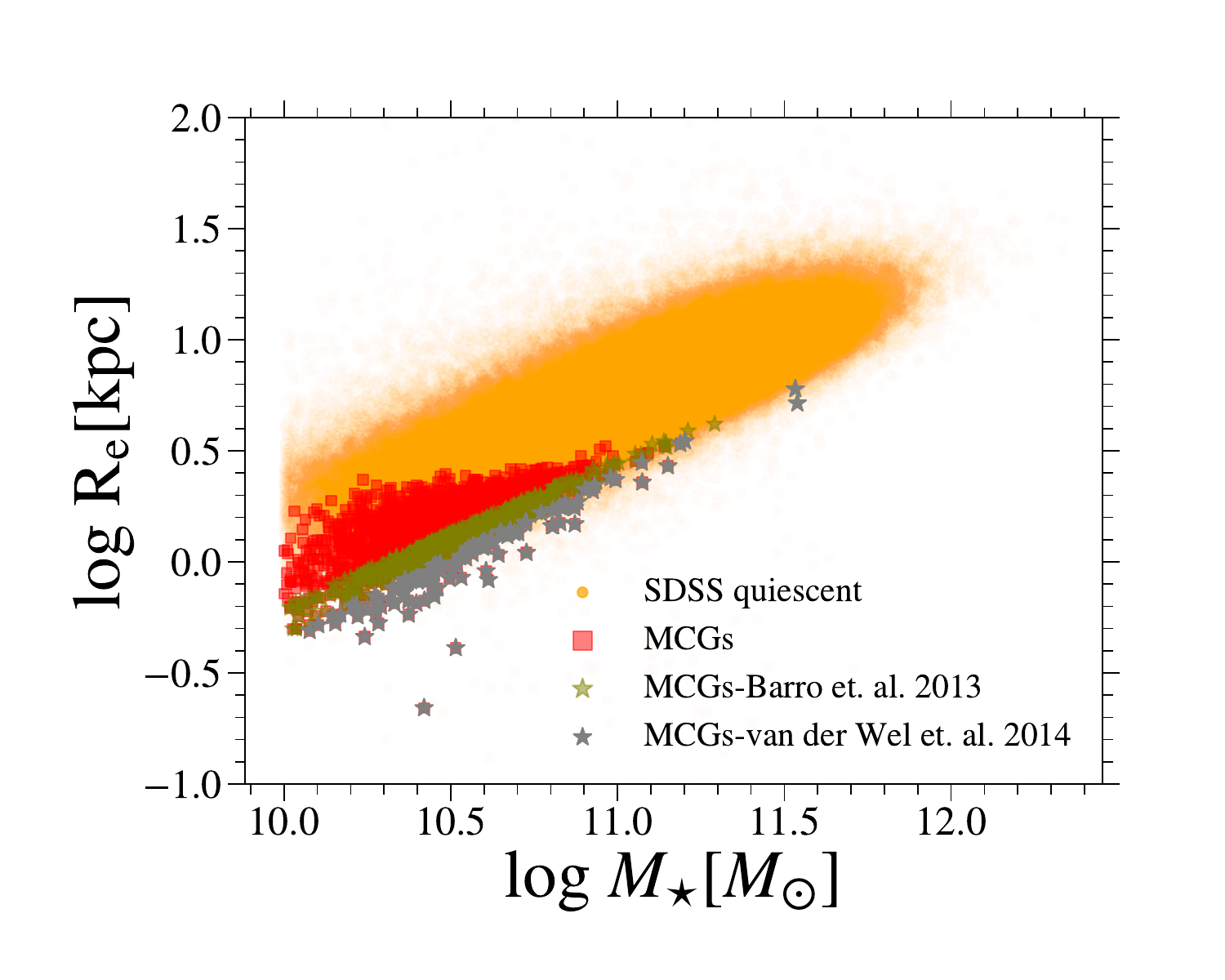}
    \caption{$\mathrm{log}\, R_{\mathrm{e}}$ as a function of   $\mathrm{log}\, M_{\star}$. SDSS quiescent galaxies are shown in orange, MCGs are shown as red squares, the subset of MCGs satisfying the \protect\cite{Barro.etal.2013} compactness criterion are shown as olive stars, and gray stars represents the subset of MCGs satisfying the \protect\cite{vanderwel.etal.2014} compactness criterion.}
    \label{fig:stellarmass_size_fig}
\end{figure}

\begin{figure*}
\centering
	\includegraphics[width=\textwidth]{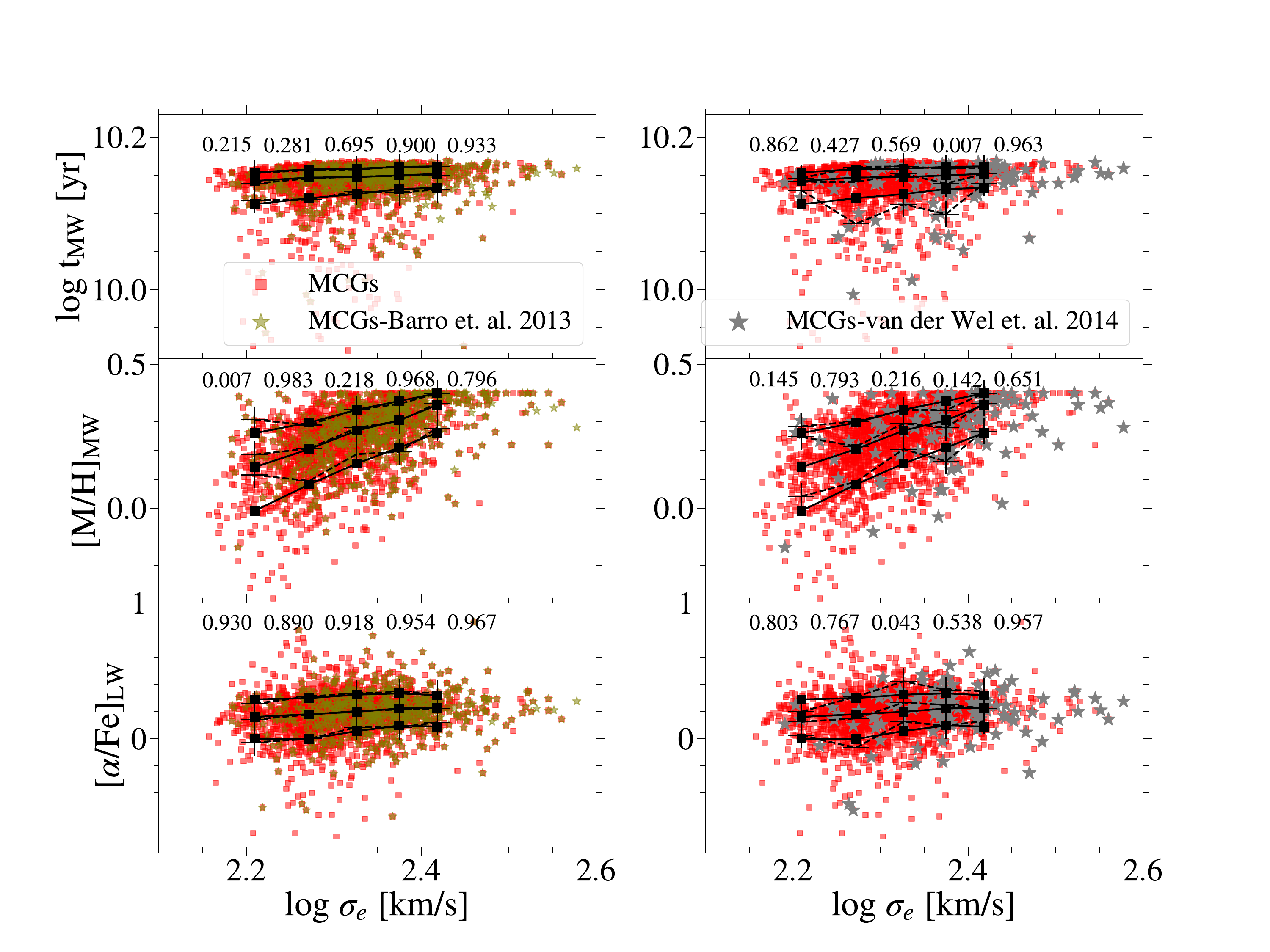}
    \caption{Comparison of stellar population properties of the original MCG sample and subsamples of  MCGs satisfying the \protect\cite{Barro.etal.2013} (left panels), and \protect\cite{vanderwel.etal.2014} (right panels) compactness criteria.}
    \label{fig:stellar_population_properties_barro_wel}
\end{figure*}

To investigate if applying an additional selection criterion based on the size-mass relation affects our results, we compared the stellar populations properties of the original MCG sample and two subsamples comprised of MCGs satisfying the compactness criterion of \cite{Barro.etal.2013} ($\log M_\star/R_\mathrm{e}^{1.5} > 10.3$) and \cite{vanderwel.etal.2014} ($R_\mathrm{e}/(M_\star/10^{11}M_\odot)^{0.75} < 2.5$\,kpc). In Fig. \ref{fig:stellarmass_size_fig} we show in red the original MCG sample, olive stars are MCGs satisfying the compactness criterion of  
 \cite{Barro.etal.2013} and gray stars are MCGs satisfying the \cite{vanderwel.etal.2014} criterion. 
In Fig. \ref{fig:stellar_population_properties_barro_wel} we show the stellar population properties of MCGs and the two subsamples. There is no statistically significant between them. 

\section{The effects of Incompleteness}
\label{sec:completeness_sdss}

As the redshift limit of our samples is $z = 0.1$, our MCG and CSG samples are incomplete for stellar masses lower than $\log M_\star \sim  10.5$. Thus, to assess the impact of incompleteness on our results, we applied a magnitude cut of $Mr < -20.4$, corresponding to the SDSS completion limit at $z \sim 0.1$.   

\begin{figure}
	\includegraphics[width=\columnwidth, trim=10 40 0 10]{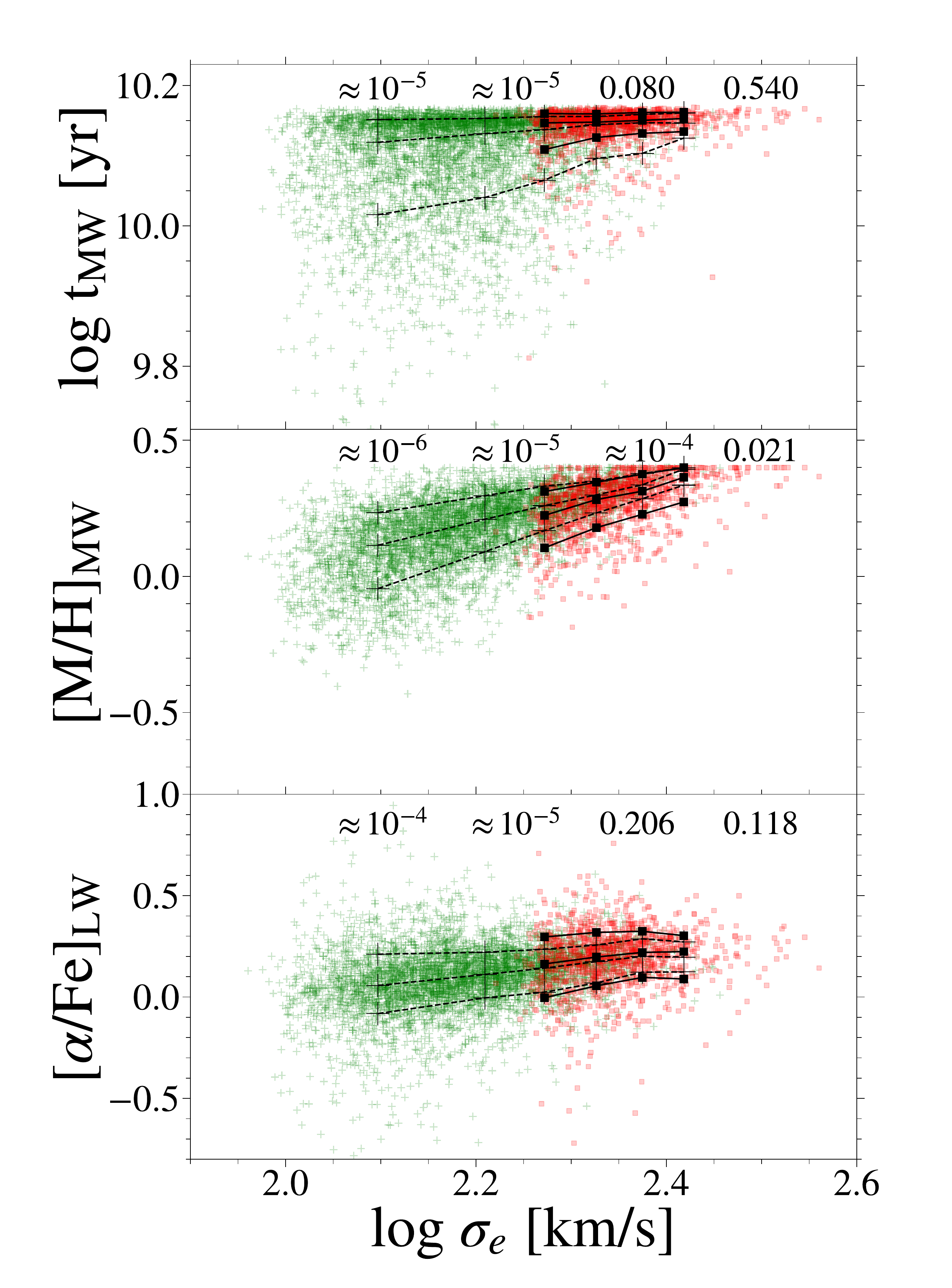}
    \caption{Stellar population properties as a function of $\sigma_\mathrm{e}$ for MCGs and CSGs with $M_r < -20.4$.}
    \label{fig:idade_metalicidade_Mr}
\end{figure}

Fig.\, \ref{fig:idade_metalicidade_Mr} shows the stellar population properties as a function of $\sigma_\mathrm{e}$ for MCGs and CSGs with $M_r < -20.4$. Applying a magnitude cut of $M_r < -20.4$ results in the loss of a large fraction of the MCGs in the $\sigma_\mathrm{e} < 175$ bin, thus we create a $\sigma_\mathrm{e} \leq 200$\,km/s bin instead.  We note that the behaviour of stellar population properties at fixed $\sigma_{\mathrm{e}}$ are the same we find in the Fig.\, \ref{fig:age_metallicity_alpha_fe_ppxf} indicating that the SDSS incompleteness does not affect our results.   

\section{Broadening correction function}
\label{sec:broadening_function}

Spectral indices are affected by the internal velocity dispersion due to a broadening of the spectral features that results in a lower strength of the indices. To correct for this effect, we built a broadening correction function based on the methodology followed by \cite{Trager.etal.1998} and \cite{delaRosa.etal.2007}. 

\begin{figure}
	\includegraphics[width=\columnwidth, trim=10 40 0 10]{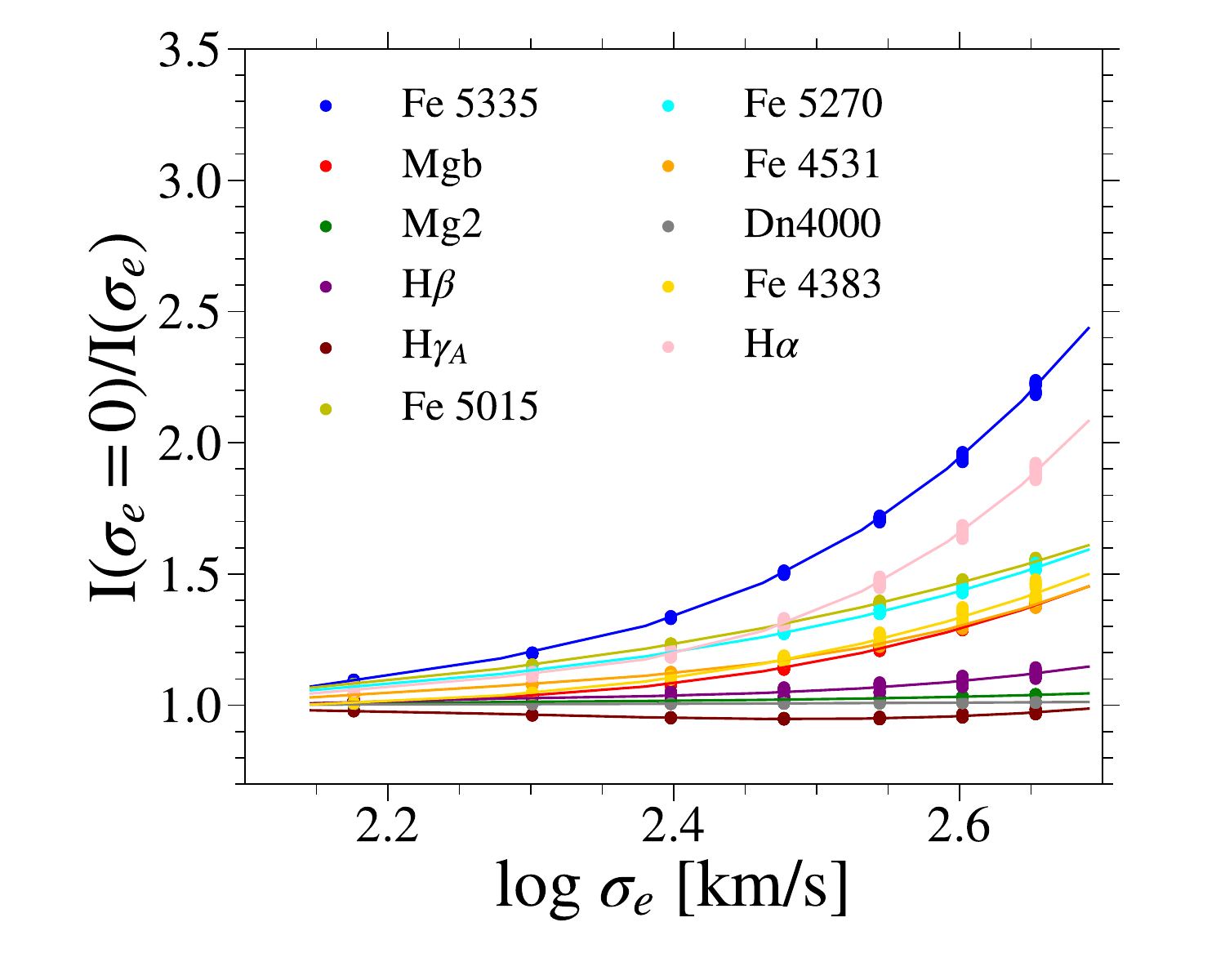}
    \caption{Ratio between the spectral indices values measured in the unbroadened ($I(\sigma_\mathrm{e} = 0$)) and broadened spectra ($I(\sigma_\mathrm{e}$)) as a function of $\sigma_\mathrm{e}$. The broadening correction function for some spectral indices are shown.}
    \label{fig:correction_function}
\end{figure}

In summary, we  selected \cite{Vazdekis.etal.2015} models with ages between $9.5$ - $13.5$ Gyr  and metallicities  between [M/H] = $-0.25$ - $+0.4$ and we convolved them with Gaussians functions with velocity dispersion values up to $400$ km/s. We used {\scshape pylick} to measure the spectral indices of the convolved spectra and then we computed the ratio between the values measured in the unbroadened ($I(\sigma_\mathrm{e} = 0$) -- equivalent to minimum $\sigma_\mathrm{e}$) and broadened spectra ($I(\sigma_\mathrm{e}$). The results are shown in shown in Fig.\,\ref{fig:correction_function}. We fitted the data with a third degree polynomial with the form: $a \cdot x^3 + b \cdot x^2 + c \cdot x + d $, where $x$ is log$_{10}  \sigma_\mathrm{e}$. The resulting coefficients for each spectral index are listed in table \ref{tab:table_coef}.

\begin{table}
	\centering
	\caption{Coefficients of the broadening correction function.}
	\label{tab:table_coef}
	\begin{tabular}{lcccr} 
		\hline
            $I(\sigma_\mathrm{e} = 0)$/$I(\sigma_\mathrm{e})$& a & b & c & d \\
		\hline
		  Fe 5335 & 7.61 & -50.00 &110.28   &  -80.53  \\
		  Mgb & 1.78 & -11.09 &23.14   &  -15.15   \\
		Mg2 & 0.30 & -2.11 & 4.94  &   -2.87 \\
		  H$\beta$ & 1.52 & -10.54 &24.39   &  -17.85  \\
		  H$\gamma_A$ & 1.24 & -8.51 & 19.41  &  -13.69  \\
		  H$\delta_A$ & 0.91 & -5.25 & 10.15  &  -5.57  \\
  	    Fe 5015 & 0.68 & -3.76 & 7.14  &   -3.70 \\
		  Fe 5270 & 1.38 & -8.55 & 18.04  &  -11.92  \\
		  Fe 4531 & 2.79 & -18.74 &42.33   &  -31.03  \\
		  Dn4000 & 0.01 & -0.07 & 0.13  &  0.91  \\
		Fe 4383 & 1.46 & -8.80 & 17.76  & -11.01   \\
            H$\alpha$ & 8.59 & -57.64 &129.32   &  -95.93  \\

	   \hline
	\end{tabular}
\end{table}

\section{Statistical analysis of the results}
\label{statistical_results}

\begin{figure*}
\centering
	\includegraphics[width=\textwidth]{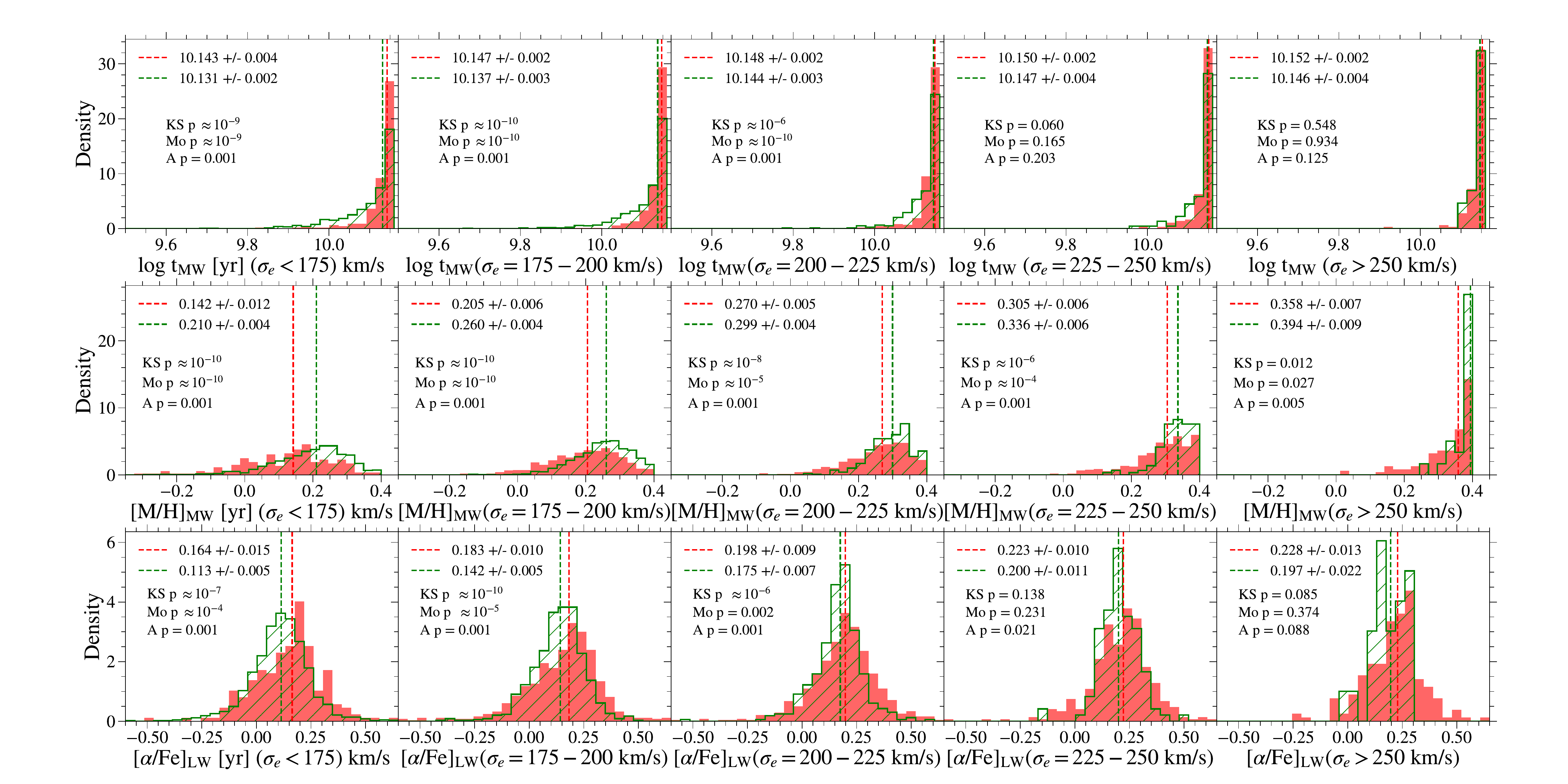}
    \caption{Distribution of Mass-weighted age, metallicity, and luminosity-weighted [$\alpha$/Fe] of MCGs (red) and of CSGs (green) separated in bins of $\sigma_\mathrm{e}$. The dashed lines show the median value,  KS p returns the KS p-value, Mo p returns the Mood’s median test , and A p returns the Anderson-Darling test.}
   \label{fig:stellar_properties_hist}
\end{figure*}

Here we show in more detail the results on the stellar population properties of MCGs and CSGs. In Fig. \ref{fig:stellar_properties_hist} we show the distribution of properties in each bin of $\sigma_e$, the median values, the Mood’s median test p-value (Mo p-v), and the Anderson-Darling test (for ksamples) p-value (A p-v).

\section{Control Sample Matched in Redshift and Velocity Dispersion}
\label{sec:sigma_CSGs}

\begin{figure}
	\includegraphics[width=\columnwidth, trim=10 40 0 10]{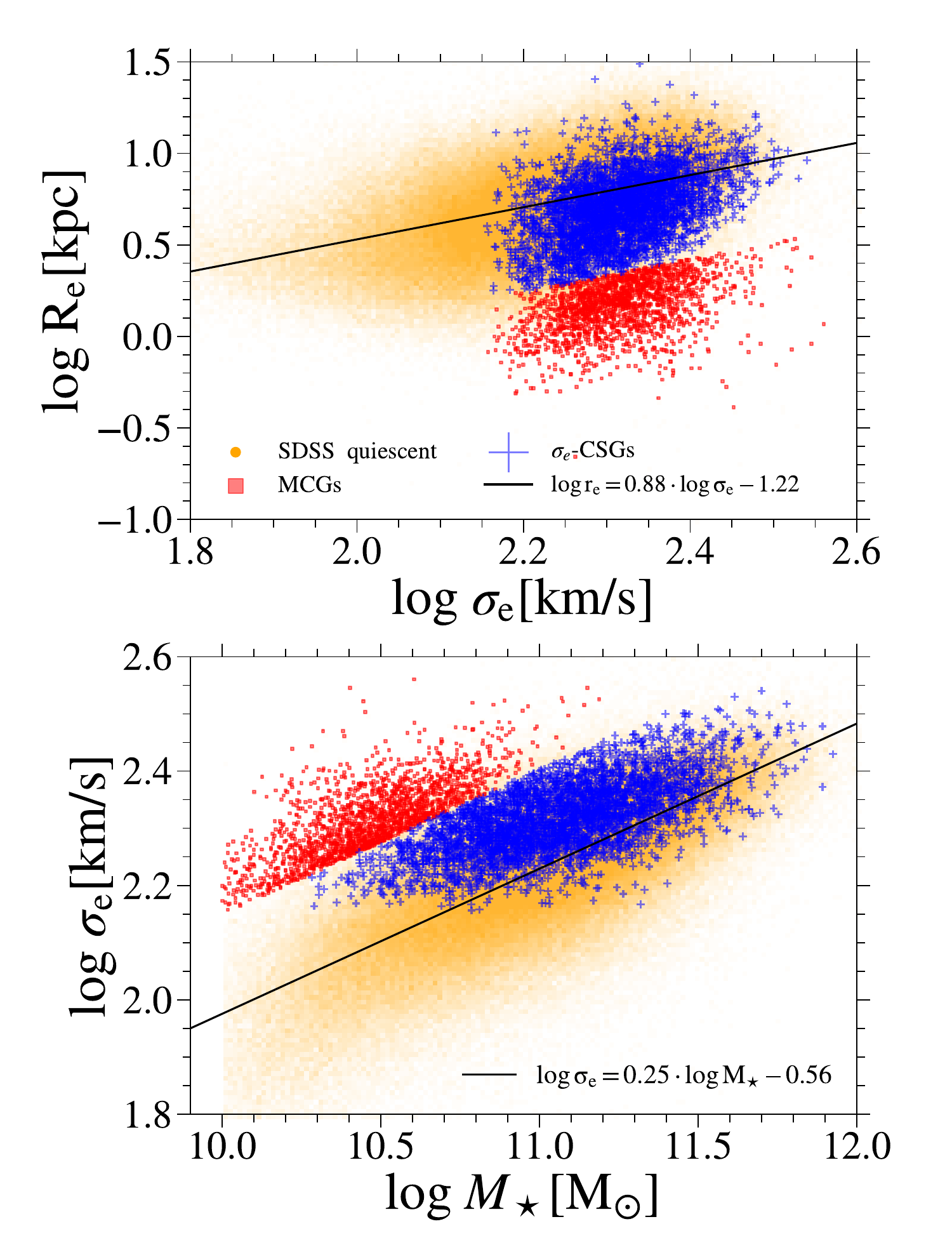}
    \caption{Similar to Fig.\, \ref{fig:final_sample}. $\sigma_\mathrm{e}$-CSGs are shown in blue.}
    \label{fig:amostra_sigma_z}
\end{figure}

Galaxies with $\sigma_\mathrm{e} \gtrsim 250$\,km/s are rare, and even with a CSG sample three times the size of the MCG sample, there were too few CSGs at high-$\sigma_\mathrm{e}$. With the goal of exploring how the trends discussed in Sec.\,\ref{synthesis} extend to higher $\sigma_\mathrm{e}$, we built a control sample of quiescent galaxies matched in $z$ and $\sigma_\mathrm{e}$ ($\sigma_\mathrm{e}$-CSGs). We followed the same methodology described in section \ref{sec:normal_quiescent_sample} to select the $\sigma_\mathrm{e}$-CSGs, although this time our $\sigma_\mathrm{e}$-CSGs has twice the size of the MCG sample. In Fig.\, \ref{fig:amostra_sigma_z} we show the distribution of $\sigma_\mathrm{e}$-CSGs in the $\log \sigma_\mathrm{e}$ vs. $\log R_\mathrm{e}$ and $\log M_\star$ vs. $\log \sigma_\mathrm{e}$ plots. Note that this sample is not representative of typical quiescent galaxies in SDSS as it is shifted to smaller sizes and stellar masses at fixed $\sigma_\mathrm{e}$. 

\begin{figure}
	\includegraphics[width=\columnwidth, trim=10 40 0 10]{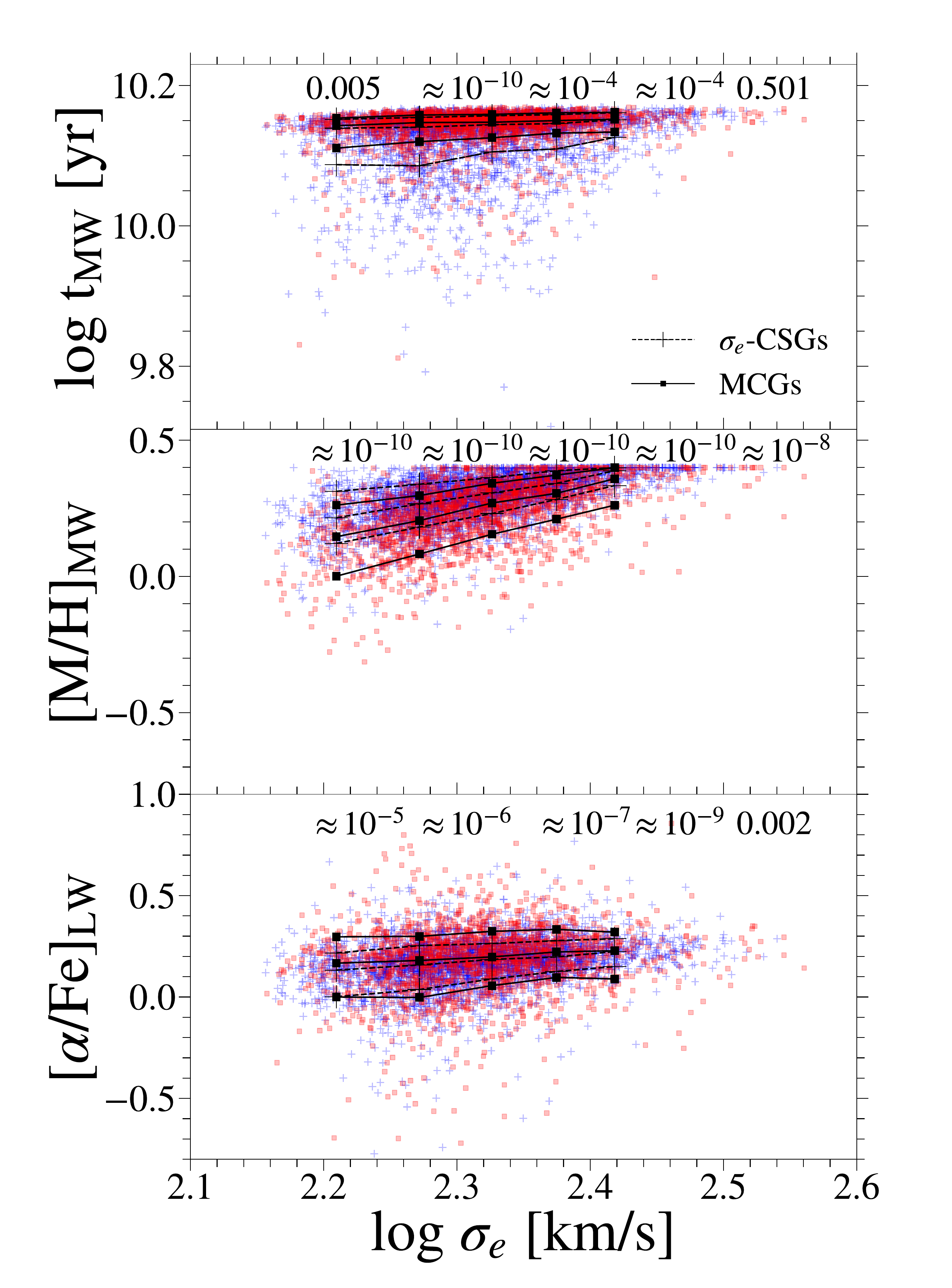}
    \caption{Similar to Fig.\, \ref{fig:age_metallicity_alpha_fe_ppxf}. The stellar population properties as a function of effective velocity dispersion are shown for MCGs (in red) and $\sigma_\mathrm{e}$-CSGs (in blue). The values in the top of each panel are the KS $p-$value tests results for each bin of $\sigma_\mathrm{e}$.}
    \label{fig:age_metallicity_alpha_fe_ppxf_sigma_z}
\end{figure}

 In Fig.\,\ref{fig:age_metallicity_alpha_fe_ppxf_sigma_z} we show the stellar population properties as a function of $\sigma_\mathrm{e}$ for MCGs and $\sigma_\mathrm{e}$-CSGs. We separated the samples into six bins of $\sigma_\mathrm{e}$, five bins with a $25$\,km/s width covering the $150-275$\,km/s interval and one bin for galaxies with $\sigma_\mathrm{e} > 275$\,km/s. At $\sigma_\mathrm{e} \lesssim 250$\,km/s the trends are similar to those in Fig.\,\ref{fig:age_metallicity_alpha_fe_ppxf}, but with the improved statistics we can better determine the behaviour of MCGs and CSGs-$\sigma_\mathrm{e}$ at $\sigma_\mathrm{e} \gtrsim 250$\,km/s: there are no statistically significant difference in ages for $\sigma_\mathrm{e} \gtrsim 275$\,km/s and differences in [$\alpha$/Fe] decrease with increasing $\sigma_\mathrm{e}$, having no statistical significance at $\sigma_\mathrm{e} \gtrsim 275$\,km/s. Differences in metallicity, however, continue being large. 

\section{Stellar surface density within 1 kpc}
\label{sec:sigma_1kpc}

\begin{figure}
	\includegraphics[width=\columnwidth, trim=10 40 0 10]{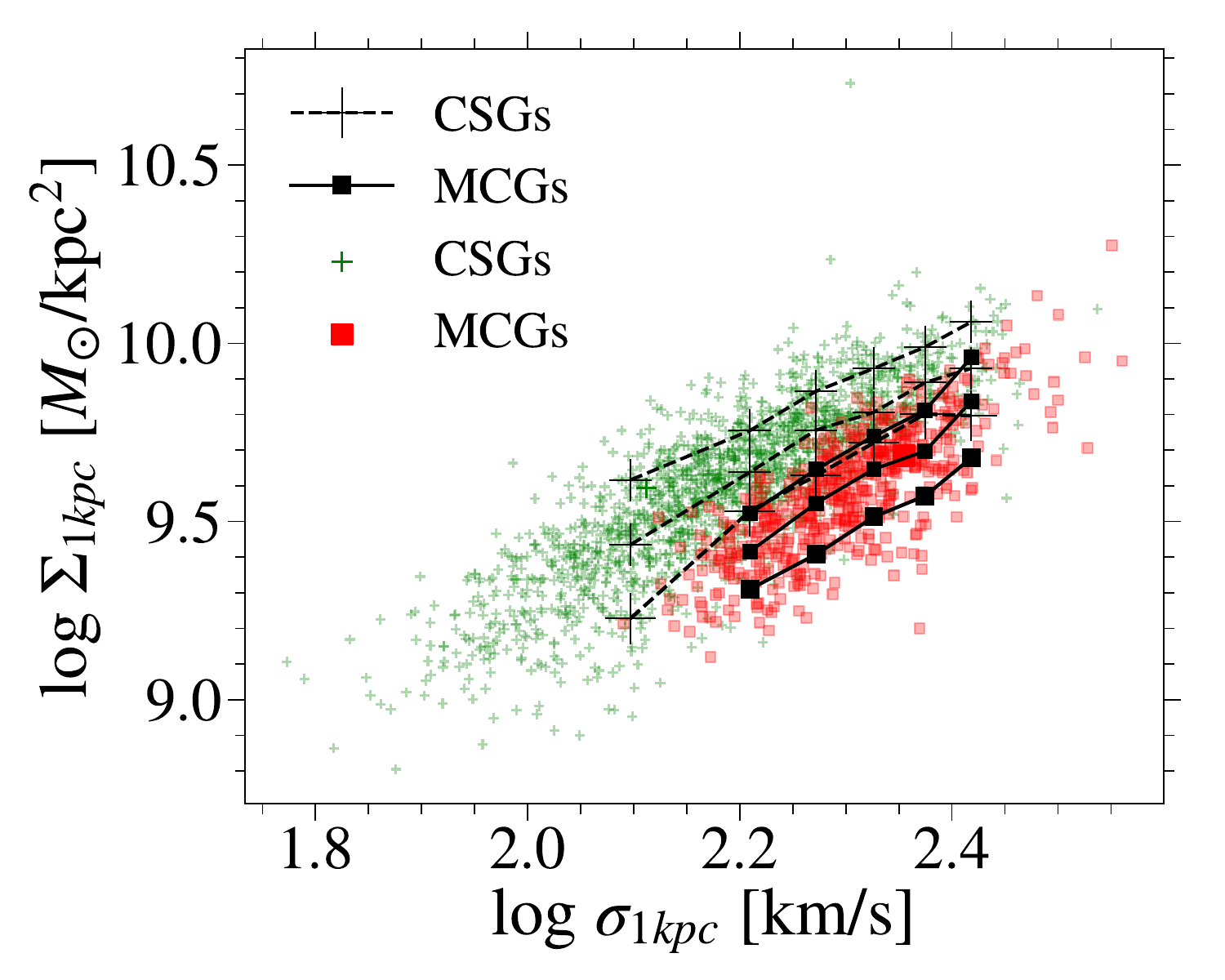}
    \caption{$\Sigma_{\mathrm{1kpc}}$ versus $\sigma_{\mathrm{1kpc}}$ for MCGs and CSGs.}
    \label{fig:Sigma_sigma_1kpc}
\end{figure}

In Fig. \ref{fig:massa_sigma_fiber} we showed that 
MCGs have, on average, lower $M_\mathrm{\star,fiber}$ at fixed $\sigma_\mathrm{fiber}$ compared to CSGs. A similar conclusion is reached when comparing the stellar mass surface density within 1\,kpc ($\Sigma_{\mathrm{1kpc}}$) at fixed $\sigma_{\mathrm{1kpc}}$, as shown in the Fig. \, \ref{fig:Sigma_sigma_1kpc}. $\Sigma_{\mathrm{1kpc}}$ values where extracted from the \cite{Luo.etal.2020} catalogue. They were computed by using the total light within 1 kpc of SDSS images in the $i$ band and the $M/L_i$ from \citep{fang13}. Note that this catalogue does not cover all galaxies from SDSS, containing $494$ MCGs and $1663$ CSGs. We found that at fixed $\sigma_{\mathrm{1\,kpc}}$  MCGs have, on average, lower values of  $\Sigma_{\mathrm{1kpc}}$ than CSGs, which agree with our results shown in Fig. \, \ref{fig:massa_sigma_fiber}.

\section{Negative Metallicity Gradients in MCGs}
\label{sec:mcgs_gradients}

\begin{figure}
	\includegraphics[width=\columnwidth, trim=10 40 0 10]{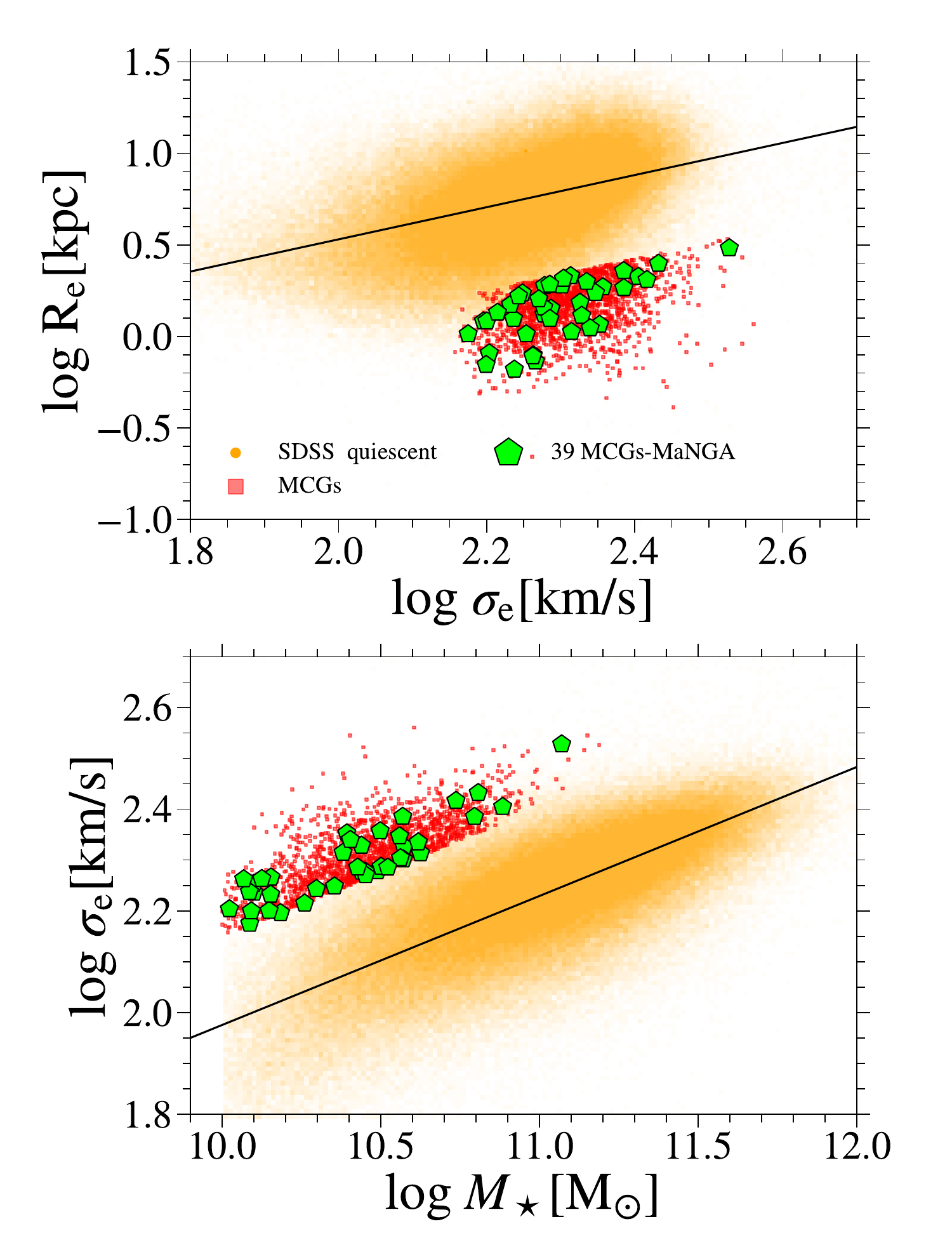}
    \caption{Similar to Fig.\,\ref{fig:final_sample}. The $39$ MCGs from the MaNGA Survey are shown as lime pentagons.}
    \label{fig:final_sample_39_MCGs}
\end{figure}

In Fig.\,\ref{fig:fibra} we see that the SDSS fiber covers $\sim 1-2\,R_\mathrm{e}$ in MCGs while covering only $\sim 0.2-1\,R_\mathrm{e}$ in CSGs. This difference needs to be taken into account when interpreting differences in metallicity between the compact and control samples, as early-type galaxies tend to exhibit significant negative gradients. To assess the impact of metallicity gradients in the measurement of metallicity in MCGs, we analysed $39$ MCGs with publicly available MaNGA data. In Fig.\,\ref{fig:final_sample_39_MCGs} we show the distribution of these galaxies in the $R_{\mathrm{e}}$-$\sigma_{\mathrm{e}}$ and $\sigma_\mathrm{e}-\log M_{\star}$ diagrams.

\begin{figure}
	\includegraphics[width=\columnwidth, trim=10 40 0 10]{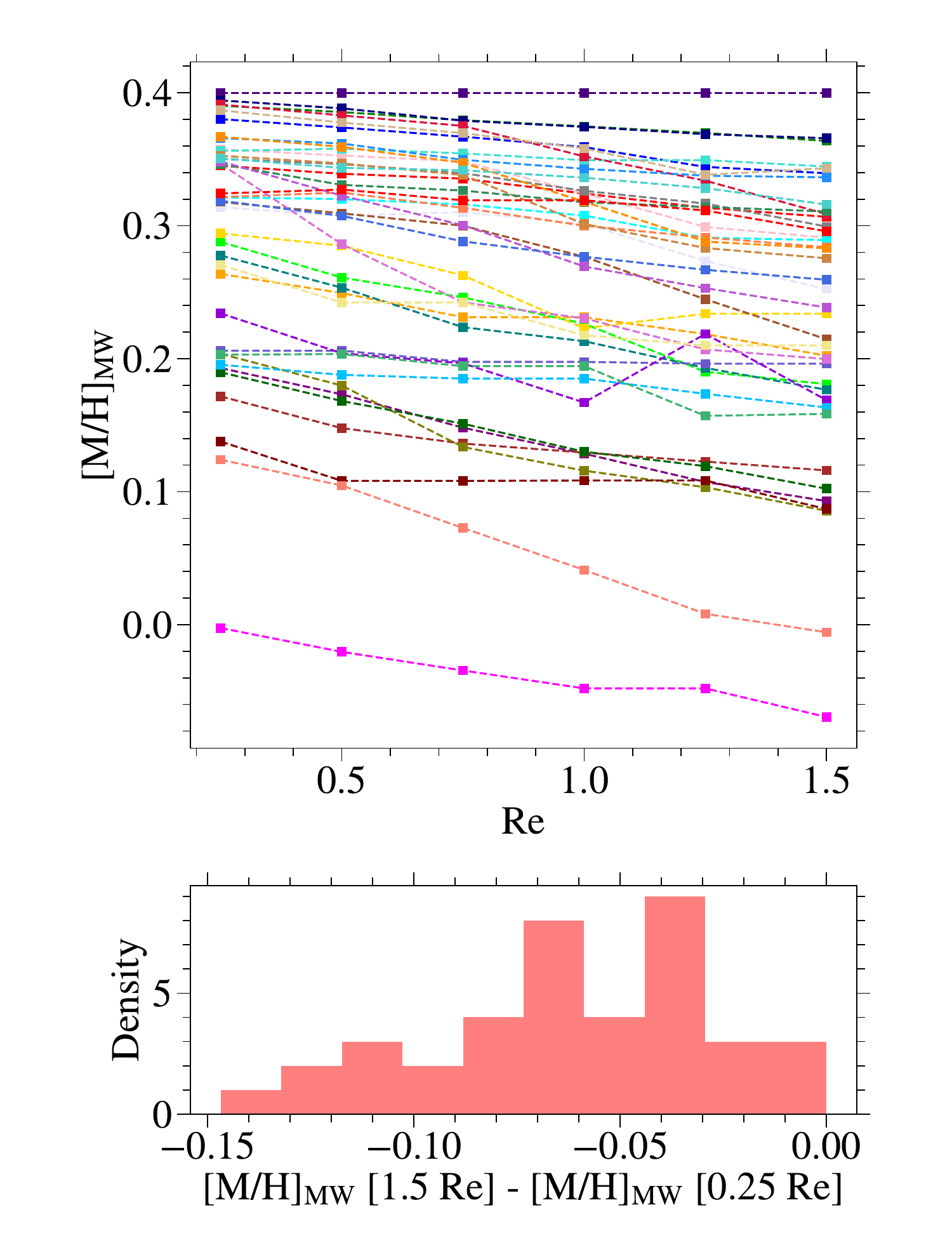}
    \caption{Stellar metallicity gradients for 39 MCGs observed as part of the MaNGA survey. Top panel: metallicity as a function of distance from the centre. Bottom panel: difference between the metallicity measured in an 0.25\,Re and 1.5\,Re aperture.}
    \label{fig:manga_grad}
\end{figure}

For each galaxy, we selected six central regions with radii of $0.25\,R_\mathrm{e}$, $0.50\,R_\mathrm{e}$, $0.75\,R_\mathrm{e}$, $1\,R_\mathrm{e}$, $1.25\,R_\mathrm{e}$, and $1.5\,R_\mathrm{e}$. We measured the mass-weighted metallicity from the integrated spectra of each region, the results are shown in Fig.\,\ref{fig:manga_grad}. We find that the metallicity decreases with radius, with a maximum variation of $0.15$ dex and a median of $0.06$ dex. Thus, the low metallicities found in SDSS MCGs could be explained, at least in part, by the larger sizes (in units of $R_\mathrm{e}$) covered by the SDSS fiber.

\bsp	
\label{lastpage}
\end{document}